\documentclass[aps,prc,twocolumn,superscriptaddress,]{revtex4-1}

\usepackage{graphicx}
\usepackage{amssymb}
\usepackage{amsthm}
\usepackage{bm}

\usepackage[utf8]{inputenc}
\usepackage[T1]{fontenc}
\usepackage{lmodern}
\usepackage{microtype}
\usepackage[english]{babel}
\usepackage[hyperindex=true,colorlinks=true]{hyperref}
\usepackage{amsmath}
\usepackage{amsfonts}
\usepackage{bm}
\usepackage{verbatim}
\usepackage{amssymb}
\usepackage{braket}
\usepackage{color}
\usepackage{wrapfig}
\usepackage{pifont}
\usepackage{amsfonts}
\usepackage{mathrsfs}
\usepackage{slashed}
\usepackage{amsmath}
\usepackage{epsfig}
\usepackage{latexsym}
\usepackage{psfrag}
\usepackage{nicefrac}
\usepackage{ragged2e}
\usepackage{etoolbox}
\usepackage{tikz}
\usepackage{pgfpages}
\usepackage{dsfont}
\usepackage{mathtools}
\usepackage{color}
\usepackage{url}

\def\orcid#1{\kern .08em\href{https://orcid.org/#1}{\includegraphics[keepaspectratio,width=0.7em]{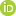}}}

\def\lnl4{N$^4$LO$_\text{lnl}$}
\def\sat{NNLO$_\text{sat}$}

\begin{document}

\title{Microscopic optical potentials for medium-mass isotopes derived at the first order of the Watson multiple scattering theory}

\author{M.\@ Vorabbi \!\orcid{0000-0002-1012-7238}}
\email[]{m.vorabbi@surrey.ac.uk}
\affiliation{Department of Physics, University of Surrey, Guildford, GU2 7XH, UK
}
\author{C.\@ Barbieri \!\orcid{0000-0001-8658-6927}}
\email[]{carlo.barbieri@unimi.it}
\affiliation{Dipartimento di Fisica, Università degli Studi di Milano, Via Celoria 16, I-20133 Milano, Italy}
\affiliation{INFN, Sezione di Milano, Via Celoria 16, I-20133 Milano, Italy}
\author{V.~Som\`a  \!\orcid{0000-0001-9386-4104 }}
\email[]{vittorio.soma@cea.fr}
\affiliation{IRFU, CEA, Universit\'e Paris-Saclay, 91191 Gif-sur-Yvette, France}
\author{P.\@ Finelli \!\orcid{0000-0002-9958-993X}}
\email[]{paolo.finelli@bo.infn.it}
\affiliation{Dipartimento di Fisica e Astronomia, Università degli Studi di Bologna and INFN, Sezione di Bologna, Via Irnerio 46, I-40126 Bologna, Italy}
\author{C.\@ Giusti \!\orcid{ 0000-0003-1901-0885}}
\email[]{carlotta.giusti@pv.infn.it}
\affiliation{INFN, Sezione di Pavia, Via Bassi 6, I-27100 Pavia, Italy}

\date{\today}

\begin{abstract}

We perform a first-principle calculation of optical potentials for nucleon elastic scattering off medium-mass isotopes. 
Fully based on a saturating chiral Hamiltonian, the optical potentials are derived by folding nuclear density distributions computed with \emph{ab initio} self-consistent Green's function theory with a nucleon-nucleon $t$ matrix computed with a consistent chiral interaction. 
The dependence on the folding interaction as well as the convergence of the target densities are investigated.
Numerical results are presented and discussed for differential cross sections and analyzing powers, with focus on elastic proton scattering off Calcium and Nickel isotopes.
Our optical potentials generally show a remarkable agreement with the available experimental data for laboratory energies in the range 65-200 MeV. 
We study the evolution of the scattering observables with increasing proton-neutron asymmetry by computing theoretical predictions of the cross section and analyzing power over the Calcium and Nickel isotopic chains.
\end{abstract}

\maketitle


\section{Introduction}

The nuclear optical potential (OP) represents a successful tool to describe the elastic nucleon-nucleus ($NA$) interaction.
It can also be extended to inelastic scattering and to the calculation of reaction cross sections for a wide variety of nuclear reactions.
The OP reduces the complexity of the quantum many-body scattering to that of a one-body Schr\"odinger equation  that is tractable across a large range of energies, target isotopes, and reaction channels \cite{Hebborn:2022vzm}. 
The basic idea is to describe the interaction between the projectile and the target with an effective complex and energy-dependent potential \cite{FESHBACH1958357, hodgson1963}.
The imaginary part accounts for the flux lost from the elastic channel to open inelastic and
reaction channels, while the energy dependence and nonlocalities account for the underlying many-nucleon dynamics.

Several ways to devise working OPs have been proposed over the years, using both phenomenological and microscopic approaches.
Phenomenological potentials typically assume an analytical form for the real and imaginary parts that characterizes the shape of the nuclear density distribution and depends on the scattering energy and the target mass number through adjustable parameters fitted to elastic $NA$ scattering data. These models are quite successful in the description of elastic scattering data and are usually adopted in the interpretation of  experimental data~\cite{Hebborn:2022vzm,KONING2003231}.

Microscopic OPs aim at avoiding the fitting procedure and are better grounded in underlying quantum mechanics. Unfortunately,  microscopic calculations imply solving the full many-body problem for the incident nucleon and all the nucleons of the target nucleus, which is a tremendous task, often beyond available computing capabilities. Clearly, some approximations are needed to reduce the problem to a tractable form and the OP depends on the reliability of such choices. As a consequence, one would expect a microscopic OP to be less accurate in describing elastic $NA$ scattering data than its phenomenological counterparts.
On the other hand, microscopic OPs are more likely to have a greater predictive power when applied to situations for which experimental information is not yet available.

From a formal point of view, the solution of the Schr\"odinger equation with an OP is equivalent to the projection of the full many-body reaction state on the subspace of the target and projectile ground states. Thus, even if the resulting OP usually has a simple structure, it has a very complicated derivation. The theoretical framework for doing so has been the focus of early pioneering works, including those of Kerman, {\it et al.}~\cite{Kerman:1959fr}, Feshbach~\cite{FESHBACH1958357,FESHBACH1962287}, Picklesimer {\it et al.} \cite{PhysRevC.30.1861}, and Watson \cite{PhysRev.105.1388} among many others. Most of these ideas, however, did not go beyond being formal developments for a long time because of the unavailability of adequate computational power.  Yet, present-day resources allow attempting at an \emph{ab initio} description of the OP, that is starting from microscopic two-nucleon ($NN$) and three-nucleon ($3N$) interactions and with approximations and theoretical uncertainties estimated and reduced in a systematic fashion~\cite{Ekstrom:2022yea}.  
This is particularly important for nuclei away from stability, whose study represents a frontier in nuclear science over the coming years and which will be probed at new rare-isotope beam facilities worldwide~\cite{Hebborn:2022vzm}.
Several groups have begun pursuing the calculation of microscopic OPs in recent years, following different routes. 
Among them, let us mention the self-consistent Green's function (SCGF) approach \cite{Barbieri2005NAscatt,Waldecker2011dom,Idini:2016oix,PhysRevLett.123.092501}, the inversion of a Green's function based on coupled-cluster~\cite{Rotureau:2016jpf, Rotureau:2018pxk, Rotureau:2020ncy} or on no-core shell model (NCSM) calculations \cite{Burrows:2018ggt, Burrows:2020qvu, PhysRevC.106.064605, 10.3389/fphy.2022.1071971},  chiral symmetry inspired OPs \cite{PhysRevC.100.014601, PhysRevC.101.064613, PhysRevLett.127.182502}, double-folding potentials from chiral effective field theory \cite{PhysRevC.102.014622, PhysRevC.105.014606}, $g$-matrix calculations~\cite{Arellano:1990xu, Arellano:1995zz, PhysRevC.98.054617, PhysRevC.102.024611}, and the phenomenological but microscopically inspired dispersive OPs \cite{Dickhoff2019PPNP, PhysRevC.101.044303}.

The SCGF approach is particularly interesting because it also delivers accurate \emph{ab initio} predictions of ground-state observables in medium mass isotopes, including nucleon density distributions~\cite{Soma2020Front,Soma2020lnl,Soma2021moving}.  More importantly, the central quantity of the formalism, the irreducible self-energy, has been shown to provide an exact extension of Feshbach theory to include not only scattering states but also overlap functions as probed by nucleon removal processes~\cite{CAPUZZI1996147,CEDERBAUM2001169,Escher2002FeshGF}.  Hence, SCGF might offer a systematic solution of the long-standing issue of the lack of consistency between the structure and reaction theory in the interpretation of experiments. The self-energy is composed of a static part, which represents the mean-field probed by the projectile, and an energy-dependent part, that encodes virtual excitations to all possible inelastic channels. A recent SCGF benchmark provided encouraging results for low-energy neutron scattering on $^{16}$O and showed the importance of coupling to enough virtual excitations to reproduce scattering at intermediate scattering energies, in the range of 10-100~MeV~\cite{PhysRevLett.123.092501}. 

In this work, we focus on the multiple scattering theory initiated by Watson~\cite{PhysRev.105.1388,Kerman:1959fr}.
This formalism was further developed in the nineties, when it was applied using the realistic $NN$ interactions available at the time together with either phenomenological or mean-field target densities~\cite{Crespo:1990zzb, Crespo:1992zz, Arellano:1990xu, Arellano:1990zz, PhysRevC.52.1992, Elster:1996xh}.
Within Watson's theory, the OP was derived at the first order in the spectator expansion~\cite{PhysRevC.52.1992} adopting different treatments of the nuclear binding between the target nucleon and the residual nucleus.

Very recently, some of the authors began applying this approach with the goal of constructing a microscopic OP for elastic nucleon-nucleus
scattering in a fully \emph{ab initio} fashion using modern chiral interactions.
The optimum factorization of the two basic ingredients of the model, i.e. the $NN$ $t$ matrix and the nuclear density, was explored in Ref.~\cite{Vorabbi:2015nra}. 
For the $NN$ interaction in the $t$ matrix, potentials up to fourth \cite{PhysRevC.68.041001} and fifth \cite{Entem:2017gor} order in the chiral expansion were employed to study the chiral convergence of elastic proton-nucleus scattering data. 
The resulting microscopic OPs were found to perform similarly to successful phenomenological potentials when comparing to experimental data on several isotopic chains~\cite{Vorabbi:2018bav}. 
The corresponding OP model was further improved in
Ref.~\cite{PhysRevC.97.034619} by folding the $NN$ $t$ matrix with a microscopic nonlocal density computed with the \emph{ab initio} NCSM \cite{BARRETT2013131} utilizing $NN$ and $3N$ chiral forces. 
The same chiral $NN$ interaction employed to calculate the nuclear density was used to calculate the $NN$ $t$ matrix. This guarantees the consistency of the theoretical framework and improves the soundness of the numerical predictions of the OP model.
The same approach, with NCSM densities, was successively extended to describe the elastic scattering of antiprotons off several target nuclei \cite{Vorabbi:2019ciy} and of protons off nonzero spin nuclei \cite{Vorabbi:2021kho}. The role of the $3N$ interaction in the dynamic part of the OP was investigated in Ref. \cite{Vorabbi:2020cgf}, where the $3N$ interaction is approximated with a density-dependent $NN$ interaction obtained after the averaging over the Fermi sphere. In practice, in this procedure the $3N$ force acts as a medium correction of the bare $NN$ interaction used to calculate the $t$ matrix.


The use of \emph{ab initio} nuclear density matrices makes the theoretical framework more microscopic and consistent, producing OPs that are quite successful in the description of the available experimental data.  However, computations based on the NCSM method are limited to light nuclei with $A \lesssim 30$ \cite{PhysRevC.99.034321}, due to the prohibitive scaling of this approach for heavier systems.
In general, and in particular for the study of nuclei away from stability, microscopic OPs are required for a wide range of nuclei. Thus, it becomes necessary to resort to many-body approaches with better scaling with respect to the mass number that allow reaching medium-mass and heavy nuclear targets.
In the present manuscript we begin exploiting SCGF computations~\cite{Dickhoff:2004xx,Carbone:2013eqa,Barbieri2017lnp,Soma2020Front} as an input to extend the scope of the multiple scattering approach. 
Apart from the possibility of overcoming the present limitation in the atomic mass number, the SCGF approach offers some advantages. 
The energy-independent part of the self-energy is closely related to the lowest term in the multiple scattering method, hence, opening opportunities for direct comparison and cross fertilization among the two approaches to derive OPs. Furthermore, SCGF calculations could supply dressed propagators to be used in the Lippmann-Schwinger (LS) equation to simulate medium effects even within multiple scattering. This extension will be the subject of future investigations. In fact, in our opinion SCGF theory provides a natural way to compute the necessary many-body components for a microscopic OP over a wide area of the nuclear chart. In particular, its capability to handle nucleon scattering beyond the impulse approximation (IA) will be a clear breakthrough for the development of nuclear reactions.

In this work, we extend the existing approach to heavier nuclei by using nuclear densities obtained from \emph{ab initio} SCGF calculations. 
Results for cross section and analyzing power of elastic proton scattering off Ca and Ni isotopic chains are presented and discussed.
The validity of our OP approach is first checked against available experimental data.
Numerical predictions are then shown for the 
two isotopic chains, with focus on the evolution of the results with the varying neutron number. 
The main goal of this work is to 
prove that the present approach can be extended to medium-mass and heavy nuclear systems giving reliable theoretical predictions. 

The manuscript is organised as follows: In Sec. \ref{Sect2} we introduce our theoretical framework for the OP operator. Formal aspects about the calculation of the OP are given in Sec. \ref{Sect2A}, details about the microscopic SCGF densities are provided in Sec. \ref{Sect2B}, while some details about the chiral potentials can be found in Sec. \ref{Sect2C}. In Sec. \ref{Sect3} we discuss the results for the scattering observables obtained with our OP model. 
Finally, in Sec. \ref{Sect4} we draw our conclusions.


\section{Optical Potential Theory}
\label{Sect2}

In this section we outline the main steps of the derivation of our microscopic OP (details and complete derivations can be found in Refs. \cite{Vorabbi:2015nra, Vorabbi:2017rvk, Vorabbi:2018bav, Vorabbi:2019ciy,  Vorabbi:2020cgf, Vorabbi:2021kho, Riesenfeld:1956zza, Kerman:1959fr, PhysRevC.30.1861, PhysRevC.40.881, PhysRevC.41.814, PhysRevC.44.1569, PhysRevC.56.2080}).
At the heart of the standard approach to the elastic scattering of a single projectile from a target of $A$ particles is the separation of the LS equation for the transition operator $T$
\begin{equation}\label{generalscatteq}
T = V + V G_0 (E) T \, ,
\end{equation}
into two parts, {\it i.e.} an integral equation for $T$
\begin{equation}\label{firsttamp}
T = U + U G_0 (E) P T \, ,
\end{equation}
where $U$ is the optical potential operator, and an integral equation for $U$
\begin{equation}\label{optpoteq}
U = V + V G_0 (E) Q U \, .
\end{equation}
In the above equations the operator $V$ represents the external interaction between projectile and target, $G_0 (E)$ is the free propagator in the projectile plus target nucleus system,
and $P$ and $Q = \mathds{1}-P$ are projection operators that select the elastic channel. The operator $P$ is defined as
\begin{equation}
P = \frac{\ket{\Psi_0^A} \bra{\Psi_0^A}}{\braket{\Psi_0^A | \Psi_0^A}} \, ,
\end{equation}
where $\ket{\Psi_0^A}$ is the ground state wave function of the target nucleus.

With these definitions the elastic scattering transition operator can be defined as $T_{\mathrm{el}} = PTP$ and
Eq.~(\ref{firsttamp}) can be written as a one-body integral equation
\begin{equation}\label{elastictransition}
T_{\mathrm{el}} = P U P + P U P G_0 (E) T_{\mathrm{el}} \, .
\end{equation}
For the present work, we only assume the presence of two-body forces in the scattering processes, since the extension to three-body forces in the projectile-target interaction has a rather small effect, as shown in Ref. \cite{Vorabbi:2020cgf}, in particular if differential cross sections are under scrutiny. With this assumption the operator $U$ for the optical potential can be expressed as
\begin{equation}
\label{optpoteq2}
U = \sum_{i=1}^A U_i = \sum_{i=1}^A \left( v_{0i} + v_{0i} G_0 (E) Q \sum_{j=1}^A U_j \right) \, ,
\end{equation}
provided that $V = \sum_{i=1}^A v_{0i}$, where $v_{0i}$ acts between the projectile ("0") and the {\it i}th target nucleon. 
Through the introduction of an
operator $\tau_i$ which satisfies
\begin{equation}\label{firstordertau}
\tau_i = v_{0i} + v_{0i} G_0 (E) Q \tau_i \, ,
\end{equation}
we can rearrange Eq.~(\ref{optpoteq2}) as
\begin{equation}
U_i = \tau_i + \tau_i G_0 (E) Q \sum_{j\neq i} U_j \, .
\end{equation}
This rearrangement process can be continued for all $A$ target particles, so that the operator for the optical potential can be
expanded in a series of $A$ terms called the spectator expansion
\begin{equation}\label{spectatorexp}
U = \sum_{i=1}^A \tau_i + \sum_{i,j\neq i}^A \tau_{ij} + \sum_{i,j\neq i,k\neq i,j}^A \tau_{ijk} + \cdots \, .
\end{equation}
For practical calculations we will introduce the single-scattering approximation, meaning that we will only retain the first term of Eq.~(\ref{spectatorexp}).
However, the operator $\tau_i$ in Eq.~(\ref{spectatorexp}) satisfies Eq.~(\ref{firstordertau}), which is still a many-body equation because of the presence of the many-body
propagator $G_0 (E)$. Because of the projectile energies we are going to consider, it is safe, as shown in Ref. \cite{Hoffmann:1981uf}, to introduce the impulse
approximation, which can be viewed as $\tau_i \approx t_{0i}$, where the operator $t_{0i}$ can be identified with the free $NN$ $t$ matrix.
With these two approximations, the final expression for the optical potential becomes
\begin{equation}\label{op_singlescat_plu_ia}
U = \sum_{i=1}^A t_{0i} \, .
\end{equation}
In the case of the IA, one never needs to solve any integral equation for more than two particles.
As shown in Ref. \cite{Tandy:1977dw}, the IA to the Watson single-scattering term provides the best two-body approximation to a single-scattering optical potential.
This has made the IA very practical in intermediate-energy nuclear physics, with a large body of work based upon this approximation over many years.

\subsection{Definition of the optical potential}
\label{Sect2A}

In this short section we present the final expression of the optical potential in the IA. Our starting point is the elastic $(A+1)$-body transition operator of Eq.~(\ref{elastictransition}), which defines the elastic OP operator as $U_{\mathrm{el}} \equiv P U P$. 
Now we distinguish between neutrons and protons and introduce the basis $\ket{\Psi_0^A ,\,{\bm k}} \equiv \ket{\Psi_0^A} \ket{\bm k}$ to project the $U_{\mathrm{el}}$ operator with $U$ given by Eq.~(\ref{op_singlescat_plu_ia}).
Here, ${\bm k}$ and ${\bm k}^{\prime}$ denote the initial and final momenta of the projectile in the projectile-target center-of-mass frame.
With this basis, we obtain a one-body equation for the elastic transition amplitude
\begin{equation}
T_{\mathrm{el}} ({\bm k}^{\prime} , {\bm k}) = U_{\mathrm{el}} ({\bm k}^{\prime} , {\bm k}) + \int d {\bm p} \frac{U_{\mathrm{el}} ({\bm k}^{\prime} , {\bm p}) T_{\mathrm{el}} ({\bm p} , {\bm k})}{E - E (p) + i \epsilon} \, ,
\end{equation}
which requires in input the elastic optical potential $U_{\mathrm{el}}$. This can be obtained, after some manipulations (see Refs. \cite{Vorabbi:2015nra, Vorabbi:2017rvk, Vorabbi:2018bav}), evaluating the single-folding integral
\begin{equation}\label{opticalpotworkeq}
\begin{split}
U_{\mathrm{el}}^{\alpha} ({\bm q} , {\bm K} ; E) = &\sum_{N=p,n} \int d {\bm P} \; \eta ({\bm q} , {\bm K} , {\bm P}) \\
&\times t_{\alpha N} \left[ {\bm q} , \frac{1}{2} \left( \frac{A+1}{A} {\bm K} - {\bm P} \right) ; E \right] \\
&\times \rho_N \Big( {\bm P} - \frac{A-1}{2A} {\bm q} , {\bm P} + \frac{A-1}{2A} {\bm q} \Big) \, ,
\end{split}
\end{equation}
where the index $\alpha$ identifies the projectile (e.g., a proton, a neutron or an antiproton) and the new variables are defined as follows
\begin{align}
{\bm q} &\equiv {\bm k}^{\prime} - {\bm k} \, , \\
{\bm K} &\equiv \frac{1}{2} ({\bm k}^{\prime} + {\bm k}) \, .
\end{align}
Here, ${\bm q}$ represents the momentum transfer and is located along the $\hat{z}$ direction, ${\bm K}$ is the average momentum, and ${\bm P}$ is the integration variable.
The $t$ matrix is generally computed in the $NN$ frame and is not a Lorentz invariant, so it must be transformed to the $NA$ frame through
the M\o ller factor $\eta$.

When evaluated in the $NN$ center-of-mass frame, the $NN$ $t$ matrix appearing in Eq.~(\ref{opticalpotworkeq}) has the following spin structure 
\begin{equation}\label{nntmatrix}
t_{\alpha N} ({\bm \kappa}^{\prime} , {\bm \kappa}) = t_{\alpha N}^c ({\bm \kappa}^{\prime} , {\bm \kappa}) + i ({\bm \sigma} \cdot \hat{{\bm n}}) \,
t_{\alpha N}^{ls} ({\bm \kappa}^{\prime} , {\bm \kappa}) \, ,
\end{equation}
where $\hat{{\bm n}}$ is the unit vector orthogonal to the scattering plane and ${\bm \kappa}$ and ${\bm \kappa}^{\prime}$ are the initial and final relative
momenta of the two nucleons.
Inserting Eq.~(\ref{nntmatrix}) into Eq.~(\ref{opticalpotworkeq}) leads to the following spin structure of the OP
\begin{equation}
U_{\mathrm{el}}^{\alpha} ({\bm q} , {\bm K};E) = U_{\mathrm{el}}^{\alpha , c} ({\bm q} , {\bm K};E) + i ({\bm \sigma} \cdot \hat{{\bm n}}) \,
U_{\mathrm{el}}^{\alpha , ls} ({\bm q} , {\bm K};E) \, ,
\end{equation}
where $U_{\mathrm{el}}^{\alpha , c}$ and $U_{\mathrm{el}}^{\alpha , ls}$ represent the central and the spin-orbit parts of the potential, respectively.

Finally, the energy $E$ in Eq.~(\ref{opticalpotworkeq}) displays a dependence on the integration variable ${\bm P}$ and makes the calculation of the integral very complicated.
In our calculations we assume the so called fixed beam energy approximation, which consists to set $E$ at one-half the kinetic energy of the projectile in the
laboratory frame, {\it i.e.}, $E = T_{lab}/2$.

\subsection{Density matrix from SCGF}
\label{Sect2B}

We compute the density matrix using SCGF theory and its algebraic diagrammatic construction [ADC($n$)] truncation scheme at different orders $n$. We use the standard, Dyson, formulation of SCGF for closed-shell isotopes and its Gorkov extension for semi-magic open shells. While the details of our computations are covered in Refs.~\cite{Cipollone2013prl,Soma2014GkvII,Cipollone2015prc,Barbieri2017lnp,Barbieri2022Gkv3}, here we summarize the points relevant to the discussion of the present results.

In the simplest case of closed-shell nuclei, the one-body propagator is obtained as a solution of the Dyson equation
\begin{align}
g_{\alpha\beta}(\omega)=g_{\alpha\beta}^0 (\omega) + \sum_{\gamma\delta} g_{\alpha\gamma}^0(\omega) \Sigma_{\gamma\delta}^\star(\omega) g_{\delta\beta}(\omega)
\label{eq:dyson}
\end{align}
in a spherical harmonic oscillator (HO) basis of $N_{\rm max}$+$1$ shells. Here, collective Greek indices $\alpha = (n_{\alpha} , l_{\alpha} , j_{\alpha} , m_{\alpha} , \tau_{\alpha})$ 
label the quantum numbers of the basis states. Each HO state can be written as a function of momentum in the laboratory frame, ${\bm p}$, spin, $\sigma$, and isospin, $\tau$, variables as
\begin{align}
\phi_{\alpha} ({\bm p}, \sigma, \tau) \equiv \, (-i)^{l_{\alpha}}\, g_{n_{\alpha} l_{\alpha}} (p) \,
\mathcal{Y}_{j_{\alpha} m_{\alpha}}^{l_{\alpha} {\scriptstyle \frac{1}{2}}} (\hat{\bm p} , \sigma ) \, 
\chi_{{ \scriptstyle \frac{1}{2}} \tau_{\alpha}} (\tau) \, ,
\end{align}
with $g_{n_{\alpha} l_{\alpha}} (p)$ being the Hankel transform of the usual HO radial function in coordinate space and the spin-angular functions being defined as
\begin{align}
\mathcal{Y}_{j_{\alpha} m_{\alpha}}^{l_{\alpha} {\scriptstyle \frac{1}{2}}} (\hat{\bm p} , \sigma ) \equiv \sum_{m_l m_s} (l_{\alpha} m_l {\scriptstyle \frac{1}{2}} m_s | j_{\alpha} m_{\alpha}) Y_{l_{\alpha} m_l} (\hat{{\bm p}}) \chi_{{ \scriptstyle \frac{1}{2} m_s}} (\sigma) \, .
\end{align}

In Eq.~\eqref{eq:dyson},  $g_{\alpha\beta}^0(\omega)$ is the free particle propagator that describes the propagation of a nucleon subject \emph{only to the kinetic energy operator}.  The nuclear many-body correlations and dynamics are encoded in the irreducible self-energy $\Sigma_{\alpha\beta}^\star(\omega)$ that has the following general spectral representation:
\begin{align}
\Sigma^\star_{\alpha \beta}(E) ={}& \Sigma^{(\infty)}_{\alpha \beta} \! + \!  \sum_{i, j}  {\bf M}^{\dagger}_{\alpha, i} \!\! \left[ \frac{1}{E - (\textbf{K}^> \!\! + \textbf{C}) + i \eta} \right]_{i, j} \!\!\!{\bf M}_{j,\beta}
 \nonumber \\ 
 & + \sum_{r, s} {\bf N}_{\alpha, r} \! \left[ \frac{1}{E - (\textbf{K}^< \!\!+ \textbf{D}) - i \eta} \right]_{r, s} \!\!\! {\bf N}^{\dagger}_{s,\beta} \, ,
 \label{eq:Sigma_ho}
\end{align}
where $\Sigma_{\alpha\beta}^{(\infty)}$ is the correlated and energy-independent mean field, and $\eta\rightarrow0^+$ imposes the correct boundary conditions.~
The ADC($n$) expansion provides a hierarchy of improvable approximations to the self-energy where at first order, or ADC(1), only the $\Sigma_{\alpha\beta}^{(\infty)}$ is retained and the SCGF approach reduces to a Hartree-Fock computation.  The ADC(2) scheme introduces the matrices $\bf{M}$~($\bf{N}$) that couple to the inelastic channels, expanded over intermediate state configurations with unperturbed energies $\bf{K}$. Eventually, the ADC(3) refines the approximation of $\bf{M}$~($\bf{N}$) and introduces the interaction ${\bf C}$~(${\bf D}$) among the intermediate states, hence enriching the description of virtual excitation of the target nucleus. Higher ADC($n$) orders would further improve the description of the nucleon-nucleus dynamics by accounting for even more complex inelastic channels~\cite{Barbieri2017lnp,Schirmer2018LNC}.  Note that the irreducible self energy~\eqref{eq:Sigma_ho} is itself an exact microscopic OP in the extended Feshbach formalism~\cite{CAPUZZI1996147}.

Whenever we compute open shell nuclei, the standard Dyson expansion of SCGF in Feynman diagrams is ill-defined due to degeneracies among the particle and hole spectra. These issues are resolved by adopting the Gorkov formulation of SCGF, where pairing effects are included already at the mean field level, but at the price of giving up particle number symmetry~\cite{Soma2011GkvI,Barbieri2022Gkv3}.  In practice, one considers the (zero temperature) grand canonical Hamiltonian \hbox{$\Omega\equiv H-\mu \hat{N}$}, where $\hat{N}$ is the particle number operator and the chemical potential $\mu$ is fixed to recover the correct number of particles on average. The Gorkov formalism introduces anomalous components, associated to particle number breaking, of both the propagator and the self-energy but it retains the spectral representation of Eq.~\eqref{eq:Sigma_ho} and has an analogous ADC($n$) expansion as for the Dyson case. Whenever the Gorkov approach is applied to fully closed-shell isotopes, we find that the anomalous terms become negligible and predictions reproduce very closely those obtained using the Dyson formulation.
Note that the Dyson SCGF formulation has been numerically implemented up to ADC(3) \cite{Cipollone2015prc, PhysRevC.97.054308}, while for open shells the working equations of Gorkov ADC(3) have been presented \cite{Barbieri2022Gkv3} but numerical implementation is currently available only up to ADC(2).

The density matrix is computed  from the imaginary part of the (normal) propagator, summing over the quasihole spectrum. Assuming a spherical ground state  $|\Psi^A_0\rangle$ for the target nucleus, with $J^\pi = 0^+$, we have:
\begin{equation}
\begin{split}
\rho_{\alpha\beta}  ={}& \frac 1 \pi \int_{-\infty}^{E_F} {\rm Im} \, g_{\beta\alpha}(\omega) \; d\omega   \\
  ={}&  \delta_{m_\alpha, m_\beta}  \, \delta_{j_\alpha, j_\beta} \, \delta_{l_\alpha, l_\beta} \, \delta_{\tau_\alpha, \tau_\beta} \; \rho_{n_\alpha, n_\beta}^{\tau_\alpha l_\alpha j_\alpha}
\end{split}
\end{equation}
where we put in evidence that the density matrix becomes block diagonal in angular momentum, parity and isospin.
Finally, the density matrix entering the computation of the OP, in Eq.~\eqref{opticalpotworkeq}, is  obtained by transforming from HO to momentum space:
\begin{equation}
\begin{split}
\rho ({\mathbf p}^{\prime} , {\mathbf p}) =\; & \langle \Psi^A_0 |  \psi_{\mathbf{p}^{\prime}}^\dagger \, \psi_\mathbf{p}  |\Psi^A_0\rangle   \\
   = \; &\delta_{\tau^{\prime} \tau^{} } \sum_{l j} \rho^{\tau l j} (p^{\prime} , p^{\phantom{\prime}} )    \\
&\times \sum_{m} \mathcal{Y}_{j m}^{l {\scriptstyle \frac{1}{2}}} (\hat{\bm p}^{\prime} , \sigma^{\prime})  \,
\mathcal{Y}_{j m}^{l {\scriptstyle \frac{1}{2}} \, \ast} (\hat{\bm p} , \sigma ) \, ,
\end{split}
\end{equation}
where $\psi_\mathbf{p}^\dagger$ ($\psi_\mathbf{p}$) are the creation (annihilation) operators for a nucleon with quantum numbers $\mathbf{p} \equiv ({\bm p}, \sigma , \tau )$ and 
\begin{equation}
\rho^{\tau l j} (p^{\prime} , p^{\phantom{\prime}} ) = \sum_{n_{\alpha} n_{\beta}}  g_{n_{\alpha} l} (p^{\prime}) \, \rho_{n_\alpha, n_\beta}^{\tau l j} \, g_{n_{\beta} l} (p) \, .
\label{eq:HO_expansion}
\end{equation}

Although we express the density matrix in momentum space, our SCGF computations are performed in a HO space limited by $N_{\rm max}$. This implies that the expansion in Eq.~\eqref{eq:HO_expansion} yields an unphysical gaussian tail at large momenta. The correct exponential asymptotic behaviour is simply recovered by repeating the diagonalization of the Dyson equation in a much larger model space where the kinetic energy in $g_{\alpha,\beta}^0(\omega)$ was computed with up to 300 HO shells, while the self-energy~\eqref{eq:Sigma_ho} (and the many-body dynamics it describes) remains truncated at $N_{\rm max}$.  We refer to this correction as ``full $\hat{T}$ space'' in the following.

\subsection{Microscopic interaction}
\label{Sect2C}

We base our computations on two- plus three-nucleon forces derived within chiral effective field theory (EFT).
In this framework, interactions between nucleons are modelled via the exchange of pions and short-range contact operators, with all possible interaction terms respecting the symmetries of quantum chromodynamics.
All these terms are organized according to their importance via a so-called power counting, which in principle provides a systematic way of truncating the EFT expansion and assessing the corresponding theoretical uncertainties~\cite{Epelbaum2009, Hammer2020}.
Modern $NN$ potentials have been constructed up to the fifth (N$^4$LO) or sixth (N$^5$LO) order \cite{10.3389/fphy.2020.00057} in Weinberg's power counting, while corresponding $3N$ forces have been derived up to N$^3$LO \cite{PhysRevC.91.044001}.
This approach leads to a generally good description of nuclear structure observables although it is known to have renormalizability problems~\cite{vanKolck2020}.

Several realizations of chiral EFT interactions have been made available in recent years which are capable of reproducing $NN$ phase-shifts and deuteron/triton properties with very high precision~\cite{Epelbaum:2014sza,Epelbaum:2014efa,Entem:2014msa, Entem:2017gor}. However, constraining the interactions to only few-body observables often fails to reproduce  binding energies and radii of larger nuclei simultaneously with the empirical nuclear matter saturation point~\cite{Lapoux2016prl,Soma2020lnl}. More recently, it has been found that proper saturation can be recovered if light to medium mass nuclei are also used to determine the Hamiltonian~\cite{PhysRevC.91.051301,Jiang2020DltGO}: typically, constraining to a few additional data points from Oxygen and Calcium isotopes lead to predictions of charge radii with 1-2\% accuracy for isotopes up to Sn and Pb~\cite{Arthuis2020prl,Hu2022Pb208}.

The possibility to simultaneously account for energies and radii of medium-mass nuclei motivated us to adopt the \sat{} interaction from Ref.~\cite{PhysRevC.91.051301} throughout this work. In particular, an accurate reproduction of the target radius is extremely important for a correct description of the diffraction minima in the cross section \cite{Arthuis2020prl}. 
The \sat{} was used for all the SCGF computations of the density matrix reported below, however, we also experiment in varying the $NN$ force in Eqs.~\eqref{optpoteq2} and~\eqref{firstordertau} to probe the sensitivity of our method to the interaction between projectile and target.


\section{Results}
\label{Sect3}

We computed the microscopic OPs for elastic nucleon scattering off Calcium and Nickel isotopes and report results for the cross section and analyzing power with nucleon incident energies between 65 and 200 MeV. This is a bit below the lower end of the range of applicability of the multiple scattering theory but still compatible with the assumptions made.
In fact, the impulse approximation adopted in the derivation of our OP works best at 200~MeV and above, but it has been found to be reasonable even for laboratory energies down to 100~MeV~\cite{Vorabbi:2021kho}.
Following Ref.~\cite{Soma2020lnl}, the density matrix of each target was obtained from SCGF theory truncating the HO basis $\{\alpha\}$ to \hbox{$N_\alpha=2\,n_\alpha+l_\alpha \leq N_{\rm max}$}. All kinetic energy and $NN$ matrix elements were included to cover the entire model space, while the $3N$ interaction was truncated to configurations with $N_\alpha+N_\beta+N_\gamma \leq E_{\rm 3max}$ because of computational resources. Unless stated otherwise, we used $N_{\rm max}$=13, $E_{\rm 3max}$=16 and an oscillator frequency of $\hbar\Omega$=14~MeV as the parameters that guarantee the best convergence of radii with respect to the model space~\cite{Soma2020lnl}. Note that the density matrices are eventually expressed in momentum space and the OP is  computed in the complete one-nucleon scattering space, as explained in Secs.~\ref{Sect2A} and~\ref{Sect2B}.

We first investigate the dependence of the scattering observables on the details of the \emph{ab initio} SCGF computations and on the chiral potential used in the $NN$ $t$ matrix. Sec. \ref{Sec3B} tests the reliability of our OPs in comparison with available experimental data and Sec. \ref{Sec3C} provides predictions for the evolution of scattering observables as a function of the neutron-to-proton asymmetry.

\subsection{Convergence and accuracy}
\label{Sec3A}

\begin{figure}[t]
\includegraphics[width= \columnwidth]{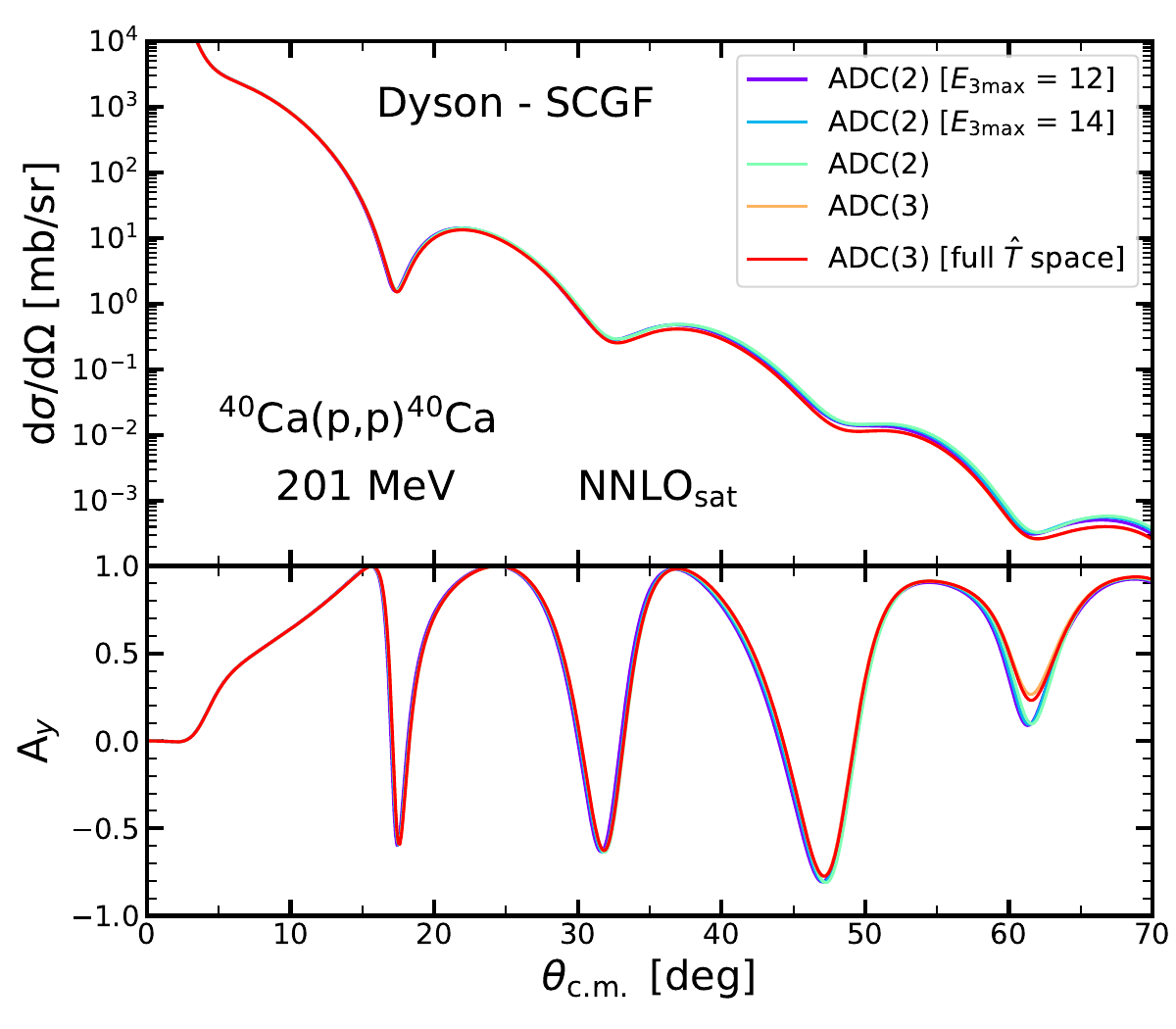}
\caption{\label{fig:Conv_Ca40}
Differential cross section
(top panel) and analyzing power
(bottom panel) as a function of the c.m. scattering angle $\theta_{\rm c.m.}$ for elastic proton scattering off $^{40}$Ca at a  laboratory energy of 201 MeV. All calculations are performed with \sat{} \cite{PhysRevC.91.051301} for both the nuclear density and the $NN$ $t$ matrix. The different curves represent various truncations employed in computing the~\emph{ab initio} density matrix. $E_{\rm 3max}$=12, 14 are a reduced truncation of $3N$ forces, all the remaining parameters being unchanged. All other curves are for $E_{\rm 3max}$=16. The ADC(3) is an improvement on the coupling to inelastic channels and ``full $\hat{T}$ space'' corrects the Gaussian HO tail in the density matrix (see discussion below Eq.~\eqref{eq:HO_expansion}).
All curves employ the Dyson formulation of SCGF theory.}
\end{figure}

\begin{figure}[h!]
\includegraphics[width= \columnwidth]{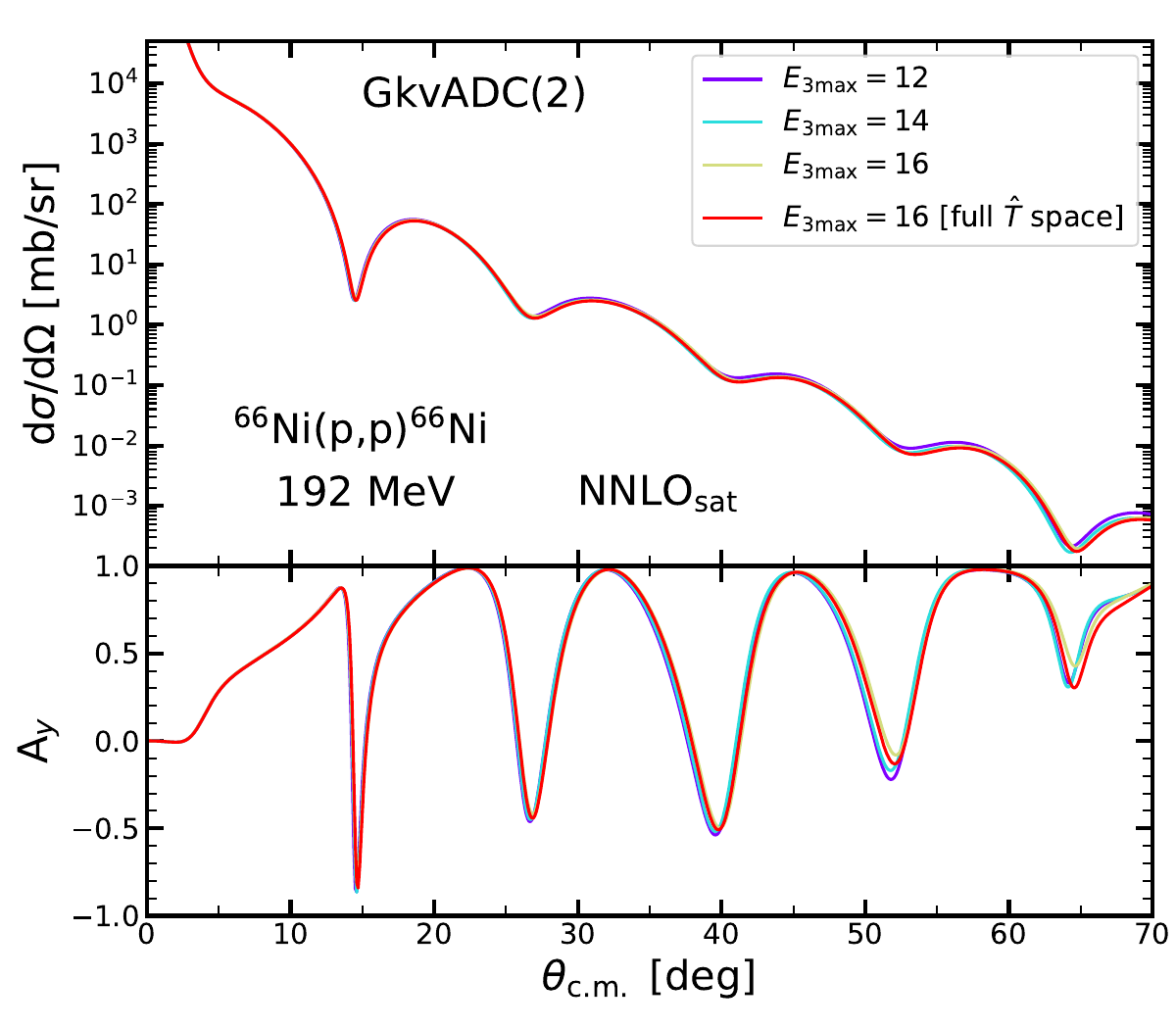}%
\caption{\label{fig:Conv_Ni66} 
Differential cross section (top panel) and analyzing power (bottom panel) for elastic proton scattering off $^{\rm 66}$Ni at a laboratory energy of 192 MeV.
Calculations are performed with \sat{} for both the $NN$ $t$ matrix and the nuclear density, obtained from GkvADC(2) SCGF calculations. The different curves represent various truncations in the \emph{ab initio} density matrix, with respect to the model space available to $3N$ forces and the kinetic energy, similarly to Fig.~\ref{fig:Conv_Ca40}.}
\end{figure}

Previous works show that the microscopic nucleon density profiles converge quickly with respect to the truncation of many-body physics and charge distributions are already accurate at the ADC(2) level for SCGF approaches~\cite{Barbieri2019Ar40nu,Lecluse2017Si34}.
Since our OPs also depend on the off diagonal part of the density matrix we perform additional ADC(3) computations 
to confirm that these findings remain valid in our case. 
Figure~\ref{fig:Conv_Ca40} demonstrates the convergence with respect to several details on the computation of the density matrix. We show the differential cross section 
(${\rm d}\sigma/{\rm d} \Omega$) and analyzing power ($A_y$) as a function of the center of mass (c.m.) scattering angle $\theta_{\rm c.m.}$ for elastic proton scattering off $^{40}$Ca at a laboratory energy of 201 MeV. 
The first three curves refer to a Dyson-ADC(2) computations with varying truncations of the $3N$ forces in the range $E_{\rm 3max}$=12-16.
The remaining curves are for ADC(3), with the usual HO truncation (at $N_{\rm max}$=13) but in one case we extend the kinetic energy to the infinite model space limit to correct for asymptotic tail in the nucleon density distribution, $\rho(r)$.
The differences between all curves shown in Fig. \ref{fig:Conv_Ca40} are very small and practically negligible both for cross section and analyzing power.
Only a very slight difference between the ADC(2) and ADC(3) curves may be noticed for the largest angles shown in the figure. 

Figure~\ref{fig:Conv_Ni66} reports analogous results for the more difficult case of an open shell isotope. Here, the differential cross section and analyzing power are shown for proton scattering off $^{66}$Ni at a laboratory energy of 192 MeV. Computations are now made using the Gorkov SCGF ad second order, or GkvADC(2). Again, all the curves are practically equivalent, for both cross section and analyzing power.
Figures~\ref{fig:Conv_Ca40} and~\ref{fig:Conv_Ni66} give a clear indication that our SCGF computations of the density matrix are fully under control with respect to truncations in $3N$ forces and many-body correlations. The resulting microscopic OPs give converged result for the observables of elastic proton-nucleus scattering, with respect to these parameters.

\begin{figure}
\includegraphics[width= \columnwidth]{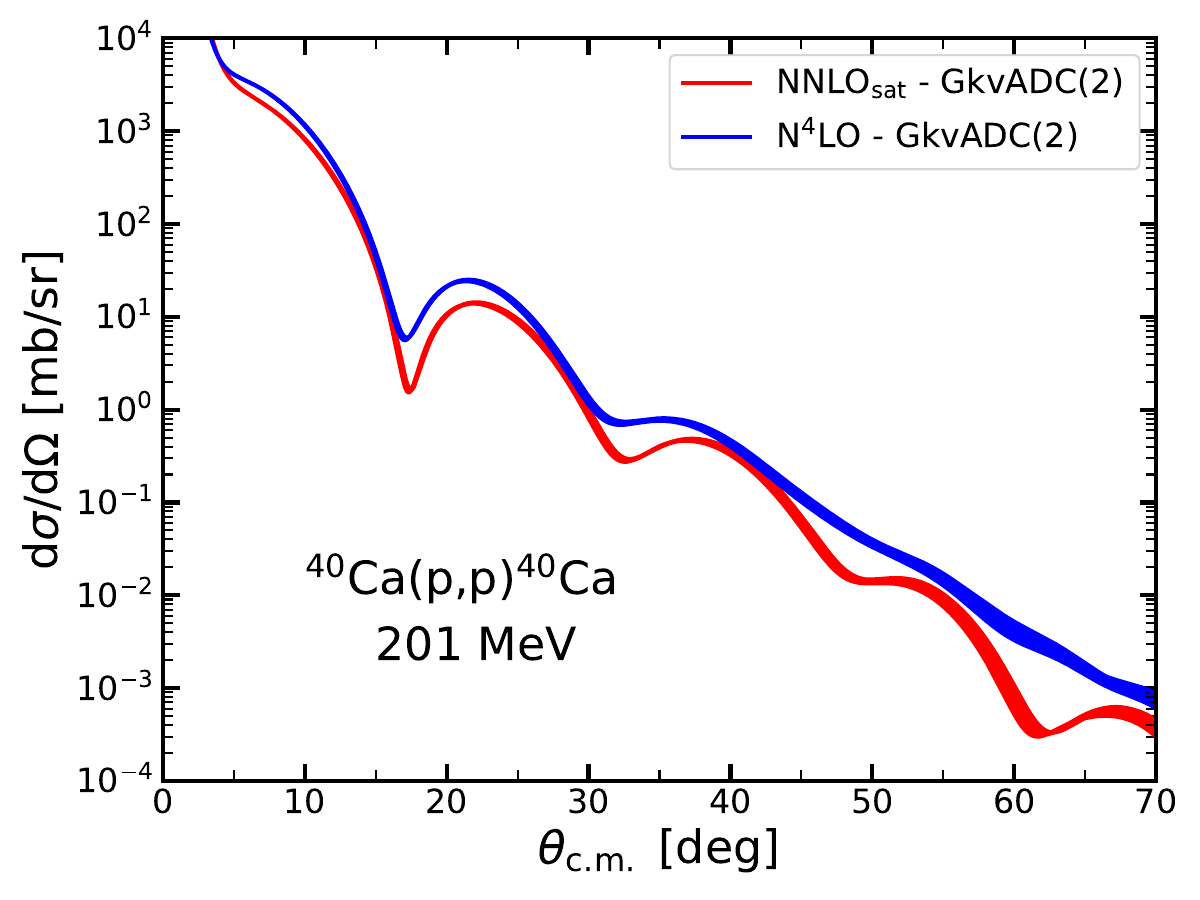}
\caption{\label{fig:Sat_vs_N4LO}
Differential cross section as a function of the c.m. scattering angle for elastic proton scattering off $^{40}$Ca at a laboratory energy of 201 MeV. Results obtained with different chiral interactions in the $NN$ $t$ matrix, \sat{} (red band) and N$^4$LO (blue band), are compared.  All results are obtained with the nuclear density matrix from GkvADC(2) SCGF calculations with \sat{}. The bands indicate the differences in the nuclear density obtained with $\hbar\Omega$=12, 14, 16, 18 and $N_{\rm max}$=11, 13.}
\end{figure}

Scattering observables can still be impacted by the choice of the model space parameters, $N_{\rm max}$ and $\hbar\Omega$, since these affect directly the computed radii. For a constant truncation, the radii typically increase monotonically when lowering the oscillator frequency, while it can be shown that they converge as a function of the HO ``effective box'' size, $L_2 =  \hbar\sqrt{2(N_{\rm max} + 7/2)/(m_N\Omega)}$, in the infrared limit~\cite{Coon2012ConvL,More2013ConvL2}. 
In practice, upper and lower limits to the converged radii can be found by choosing appropriate model spaces. For the specific case of the \sat{} interaction used in this work, we find that bounds to the converged values of radii can be found varying the model space parameters in the ranges $N_{\rm max}$=11-13 and $\hbar\Omega$=12-14~MeV~\cite{Soma2020lnl,Arthuis2020prl}. Converged \sat{} radii have tendency to overestimate the experiment by 1-2\% for Ni and light Ca isotopes~\cite{Soma2020lnl,Malbrunot2022prlNi}. However, computation for slightly larger $\hbar\Omega$ than the optimal value have been found to reproduce the experimental charge distribution more closely~\cite{Hage2016Nat_weakCa,Lecluse2017Si34,Barbieri2019Ar40nu}. 
To give a conservative estimate of the uncertainties from the model space truncation, we have recomputed $^{40}$Ca with $N_{\rm max}$=11,13 and a larger range of $\hbar\Omega$=12-18~MeV. The differential cross sections are presented in Fig.~\ref{fig:Sat_vs_N4LO} as colored bands for two separate choices of the $NN$ $t$ matrix. 
The bands are narrow and nearly negligible for forward scattering and increase only slightly with the scattering angle, without affecting the shape of the differential cross section. Thus, uncertainties in SCGF calculations of the target density do not affect significantly the theoretical predictions of our OPs. We have shown here only a few numerical examples for proton energies of about 200 MeV, but this conclusion is confirmed by corresponding results obtained over the whole range of energies to which we have applied our OPs.

\begin{figure}
\includegraphics[width= \columnwidth]{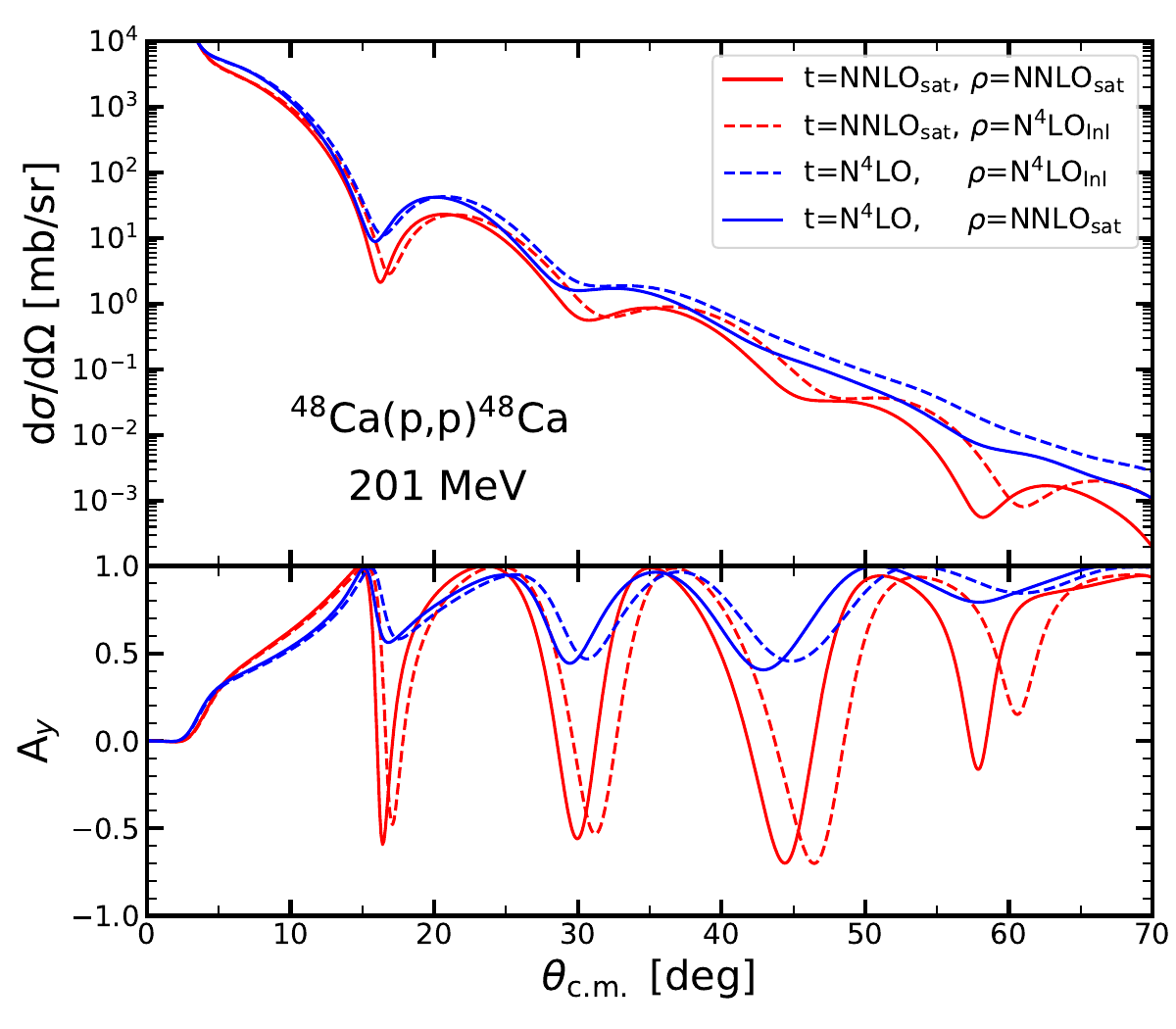}
\caption{\label{fig:Dep_vs_NNLOsat_and_N4LO}
Differential cross section (top panel) and analyzing power (bottom panel) as a function of the c.m. scattering angle $\theta_{\rm c.m.}$ for elastic proton scattering off $^{48}$Ca at a laboratory energy of 201 MeV. The solid (dashed) lines are obtained by using GkvADC(2) density matrices computed with the \sat{} (\lnl4) interactions, while red (blue) lines use the NN part of \sat{} (N$^4$LO) for the $NN$ $t$ matrix.  The SCGF computations are done with $N_{\rm max}$=13 and used an oscillator frequency of $\hbar\Omega$=14~MeV  for \sat{} and $\hbar\Omega$=20~MeV for \lnl4  as the values giving optimal convergence for nuclear radii in both cases~\cite{Soma2020lnl}.
}
\end{figure}

Next, we investigate the choice of the chiral forces in the $NN$ $t$ matrix by comparing two different models for the interaction among the projectile and the nucleons in the target. The \sat{} force is consistent with the computation of the density matrix, which is a general requirement for pursuing \emph{ab initio}, well controlled, predictions. It also has the clear advantage of reproducing correct radii. However, concerns may arise from the fact that this interaction is only fitted to low energy scattering data, up to 35~MeV, and has a poor reproduction of p-wave $NN$ phase shifts in virtue of its low order in the chiral EFT expansion. At the scattering energies we consider in the spectator model even much higher partial waves are probed. For the second model we employ the $\Lambda$=500~MeV cutoff version of the N$^4$LO interaction from Ref.~\cite{Entem:2014msa,Entem:2017gor}, which tames the above concerns.
This interaction is constructed to higher orders in chiral EFT and reproduces $NN$ partial waves up to $l$=4,5 and scattering energies up to pion-production threshold.
Fig.~\ref{fig:Sat_vs_N4LO} compares the use of the \sat{} and N$^4$LO for the elastic proton scattering cross sections off $^{40}$Ca at a laboratory energy of 201 MeV. The dependence on the $NN$ $t$ matrix is sizable. The results obtained with \sat{} and N$^4$LO differ in both shape and size, the differences can be decisive in comparison with experimental data (see Sec. \ref{Sec3B}). 
To better disentangle the implications that the choice of the Hamiltonian has on different parts of our computations we repeat a similar study for $^{48}$Ca but using an additional density matrix computed from a local-nonlocal version of the N$^4$LO Hamiltonian (here named \lnl4) as presented in Refs.~\cite{PhysRevC.97.034619,Soma2020lnl}.
Fig.~\ref{fig:Dep_vs_NNLOsat_and_N4LO} shows that the density distribution generated by \lnl4 slightly shifts the diffraction minima toward larger angles with respect to \sat{}. This is a direct consequence of the fact that this interaction systematically underestimates root mean square radii and confirms that a correct reproduction of nuclear sizes is an important constraint for describing the structure of the target. On the other hand, both cross sections and analysing powers have stronger sensitivity on the choice of the $NN$ $t$~matrix. Thus, it could be expected that refinements in the scattering approach of Sec.~\ref{Sect2A} such as accounting for three-nucleon forces in the $t$ matrix, see Ref.~\cite{Vorabbi:2020cgf}, will impact the accuracy of the present approach. 
Note that only the full-red and the dashed-blue curves in Fig.~\ref{fig:Dep_vs_NNLOsat_and_N4LO} are computed consistently with a similar interaction both in the density matrix and $NN$ $t$~channel, respectively using \sat{} or \lnl4. While consistency in the microscopic Hamiltonian remains a fundamental constraint to ensure predictive power for unstable isotopes, Figs.~\ref{fig:Sat_vs_N4LO} and~\ref{fig:Dep_vs_NNLOsat_and_N4LO} suggest that a Bayesian analysis over several chiral Hamiltonians might be important to a systematic estimation of theoretical uncertainties~\cite{Ekstrom:2022yea}.
For the reminder of this work the target density will be always computed with \sat{}, to be consistent with empirical nuclear sizes, and we neglect $3N$ forces in the $NN$ $t$ matrix.

\subsection{Comparison to experimental data}
\label{Sec3B}

\begin{figure}
\includegraphics[width= \columnwidth]{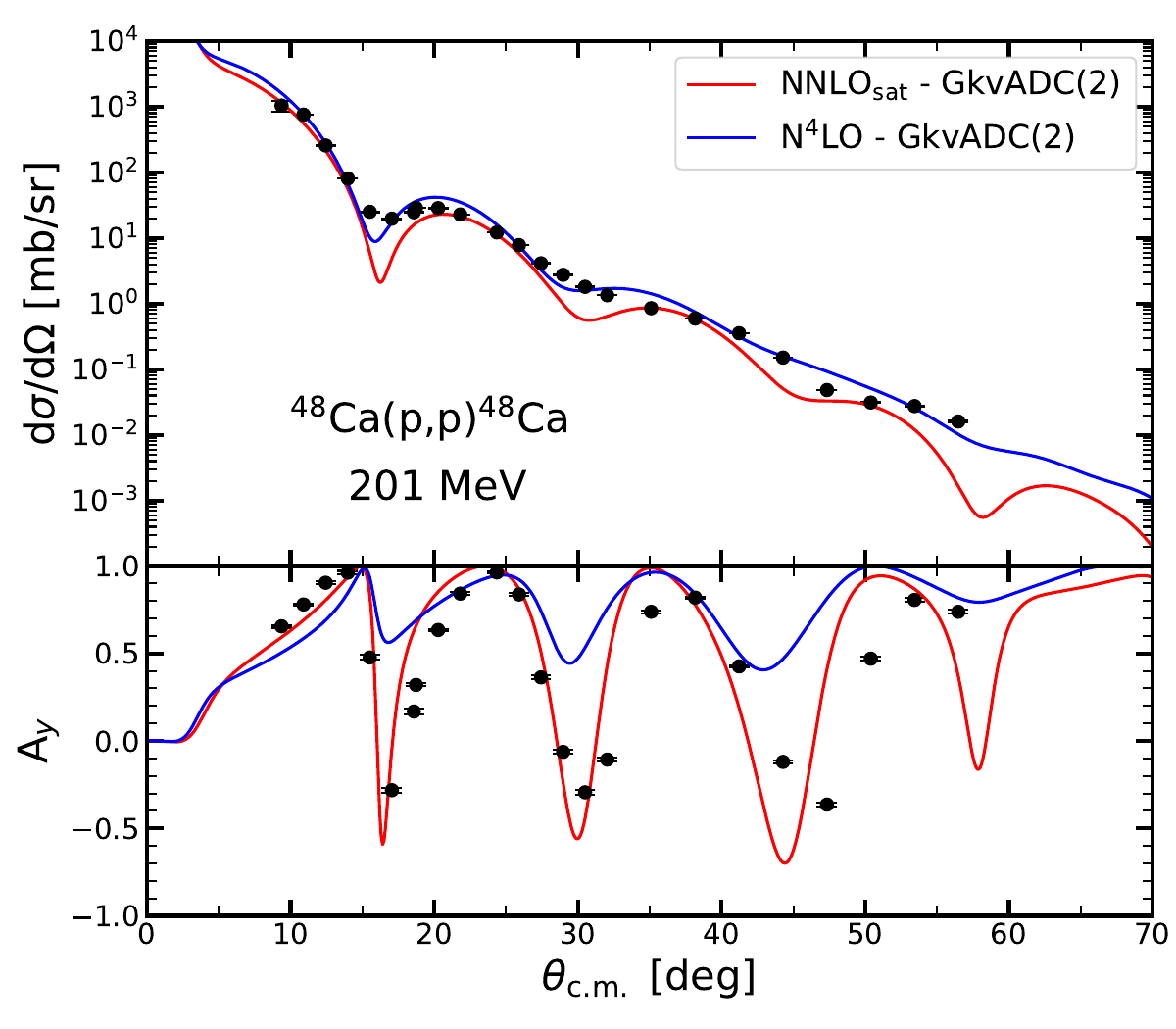}
\caption{\label{fig:Caxx_vs_exp2} 
Differential cross section (top panel) and analyzing power (bottom panel) as a function of the c.m. scattering angle $\theta_{\rm c.m.}$ for elastic proton scattering off $^{48}$Ca at a  laboratory energy of 201 MeV. Experimental data \cite{PhysRevC.49.2068} are compared with the results of microscopic OPs obtained using \sat{} (red curves) and N$^4$LO (blue curves) chiral interactions in the $NN$ $t$ matrix. In both cases, the nuclear density is obtained from GkvADC(2) SCGF calculations computed with \sat{}  with $N_{\rm max}$=13 and $\hbar\Omega$=14~MeV. }
\end{figure}

\begin{figure}
\includegraphics[width= \columnwidth]{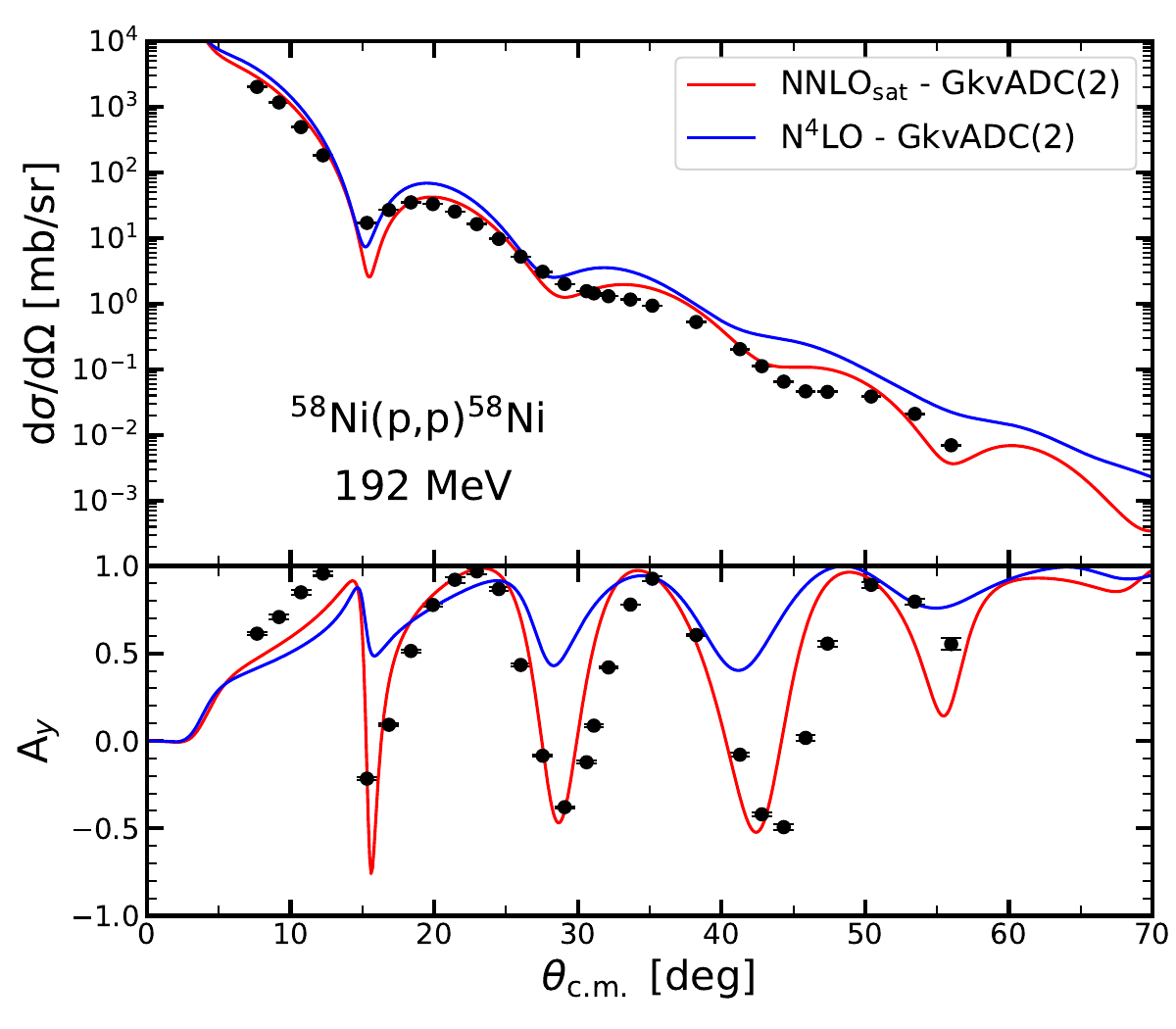}
\caption{\label{fig:Ni_vs_exp1} 
Same as in Fig.~\ref{fig:Caxx_vs_exp2} but for elastic proton scattering off $^{58}$Ni at 192~MeV. Experimental data is taken from Ref.~\cite{PhysRevC.57.1749}.}
\end{figure}

We now move to confronting the prediction of the two different $NN$ $t$ matrices and compare the results of our microscopic OPs against the available data for nucleon elastic scattering off Calcium and Nickel isotopes. 
Fig.~\ref{fig:Caxx_vs_exp2} displays the differential cross section and analyzing power as a function of the c.m. scattering angle for protons off a $^{48}$Ca target at 201~MeV laboratory energy. 
The experimental data are compared with the results obtained using the \sat{} and N$^4$LO chiral interactions in the $NN$ $t$ matrix. 
The two forces produce significant differences in both shape and size of the cross section and analyzing power. Both results give a reasonable description of the experimental cross section, although the agreement is somewhat better for \sat{}. Larger differences are found for $A_y$ where both interactions are able to describe the shape and the position of the experimental minima. However, only \sat{} reproduces their depth.
The analogous comparison for $^{58}$Ni at 192~MeV is given in Fig.~\ref{fig:Ni_vs_exp1}. It confirms that there is a significant dependence of the OPs on the chiral interactions used for the $NN$ $t$ matrix. In general, the results obtained with \sat{} give a better description of the experimental data and, in particular, a remarkable description of the experimental analyzing power. We have tested other isotopes and energies and always found confirmation of these findings, see App.~\ref{AppX}.



In Figs.~\ref{fig:Caxx_vs_exp2}, \ref{fig:Ni_vs_exp1} and in the Appendix
we have compared the results of our OPs with experimental data above 150 MeV, where the approximations adopted in our OP model are expected to be valid.
This has been already investigated and confirmed in previous work~\cite{Vorabbi:2015nra, Vorabbi:2017rvk, Vorabbi:2018bav, Vorabbi:2019ciy,  Vorabbi:2020cgf, Vorabbi:2021kho}. 
Let us remark, however, that \sat{} was constrained to much lower $NN$ scattering energies. A quick look at scattering amplitudes shows that predictions from \sat{} still compare reasonably well to the experiment up to 200~MeV although this is far from being perfect (in contrast to N$^4$LO which fits the data by construction). For a multiple scattering based approach such as the present work, it is plausible that small discrepancies with NN data due to missing higher orders in the chiral EFT expansion average out as the IA gains validity.
Even if its predicted NNLO phase shifts remain reasonable at larger energies, the good agreement on experimental analyzing powers could still be somewhat fortuitous. 
Importantly, even though the shape of the target nucleus is under control, the dependence on the interaction between the projectile and the other nucleon can be important.

\begin{figure}
\includegraphics[width= \columnwidth]{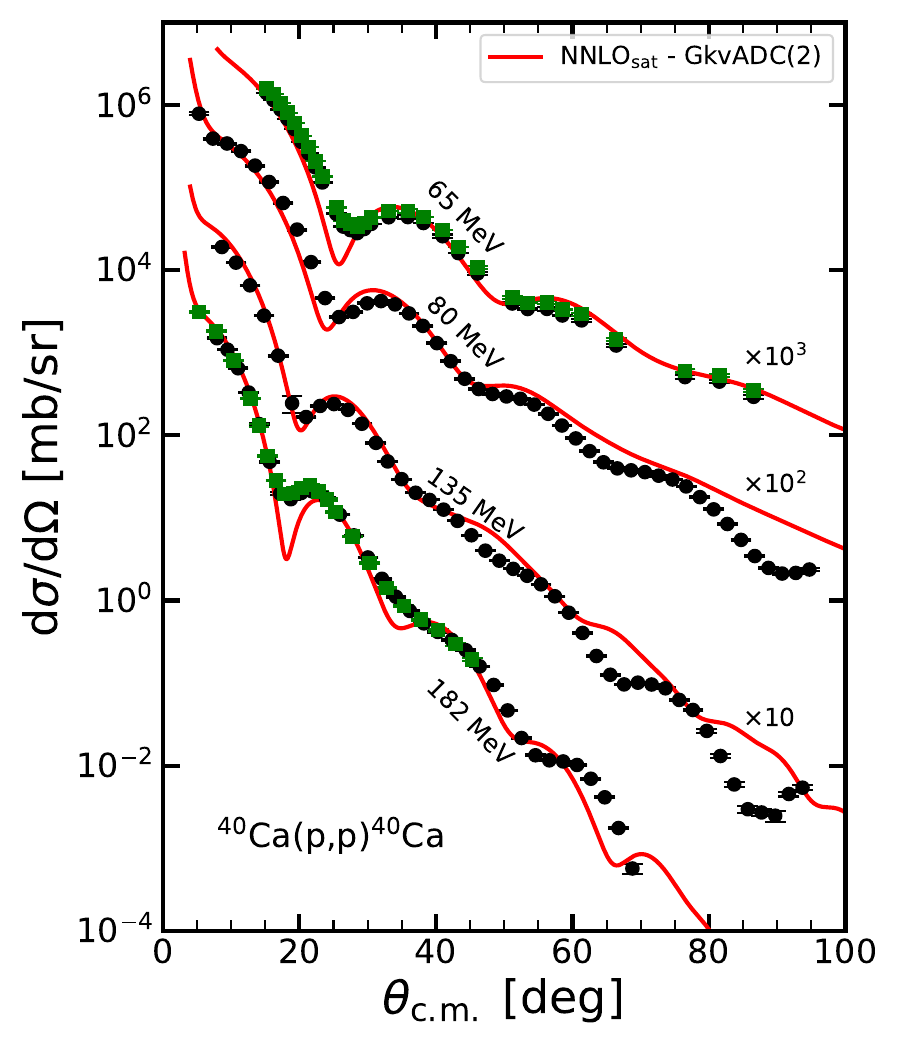}
\caption{\label{fig:Caxx_vs_exp3} 
Differential cross section as a function of the c.m. scattering angle for elastic proton scattering off $^{40}$Ca at 65, 80, 135, and 182 MeV laboratory energy. The results of the OPs obtained with GkvADC(2) densities computed from \sat{} are compared to experimental data from Ref.~\cite{PhysRevC.26.944, NORO1981189, PhysRevC.23.1023, PhysRevC.26.55, Johansson_1961}.}
\end{figure}

The comparison between the results of our OPs computed with \sat{} and the experimental differential cross sections of elastic proton scattering off $^{40}$Ca in a range of proton energies between 65 and 182 MeV is displayed in Fig.~\ref{fig:Caxx_vs_exp3}. Our OPs are able to give a reasonable description of the experimental cross section at all energies considered. The agreement gets somewhat worse for larger values of the scattering angle. We note the remarkably good agreement between our OP and the data at 65 MeV, an energy that can be considered at the limit of validity of the impulse approximation adopted in our OP model.


Overall, the agreement found between our theoretical results and the experimental data is remarkably good, and it makes our approach to the OP comparable to the other existing approaches on the market. The striking feature of our method is that allows us to compute the OP using $NN$ and $3N$ interactions as the only input, which is extremely important to maintain consistency and predictive power in our calculations.

\begin{figure}
\includegraphics[width= \columnwidth]{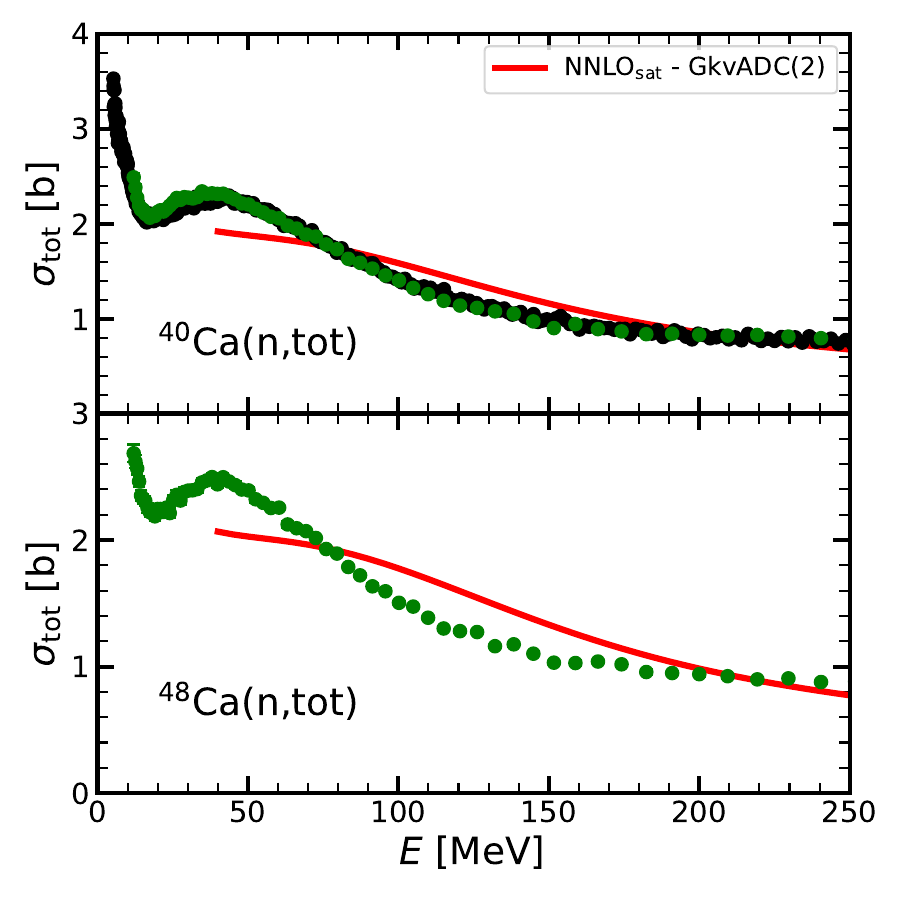}
\caption{\label{fig:Ca_total_sig} 
Total cross section for elastic neutron scattering off $^{40}$Ca (top panel) and $^{48}$Ca (bottom panel) for laboratory energies in the range 40-250~MeV. Experimental data from Ref. \cite{SHANE2010468,PhysRevC.47.237} are compared with the results of our OPs computed with \sat{}. The nuclear densities are obtained from GkvADC(2).}
\end{figure}

We now turn to predictions for the total cross sections at different energies. Fig.~\ref{fig:Ca_total_sig} shows the elastic neutron scattering cross sections off $^{40}$Ca and $^{48}$Ca for laboratory scattering energies between 40 to 250 MeV. The results of our microscopic OPs computed with the \sat{} interaction are compared with the experimental cross sections. The model adopted to derive our OPs contains several approximations and we do not expect to obtain a perfect agreement with the experimental cross section across the whole energy range considered. The main aim of this investigation is to obtain an indication of the general validity of our OP and of its approximations as a function of the neutron energy. At the highest energies considered, in the range 180-250 MeV, our OPs are able to describe the experimental total cross sections of both isotopes. This is a range for which the approximations adopted to derive our OP hold.
The adopted impulse approximation worsens gradually as the energy decreases but it does not diverge from the experiment in any abrupt way. In the range between 70-180~MeV our results overestimate the experimental cross sections of both isotopes and we find the largest disagreement with the experiment for $^{48}$Ca.
At the lowest energies considered, 40-70 MeV, the experimental cross sections for both isotopes are underpredicted. We consider these energies too low for IA to hold, without the inclusion of medium effects and any higher order correction in the spectator expansion. Yet, the overall disagreement with data is contained.
The reasonably good description of the differential cross section found in Fig.~\ref{fig:Caxx_vs_exp3} and the comparison between theoretical and experimental total cross sections suggests that improvements needed in our microscopic OPs may be relatively modest and under reach. In particular, the contributions of medium effects might reduce or even solve the discrepancies between numerical results and data found in this work.

\subsection{Calcium and Nickel isotopic chains}
\label{Sec3C}

\begin{figure}[t]
\includegraphics[width= \columnwidth]{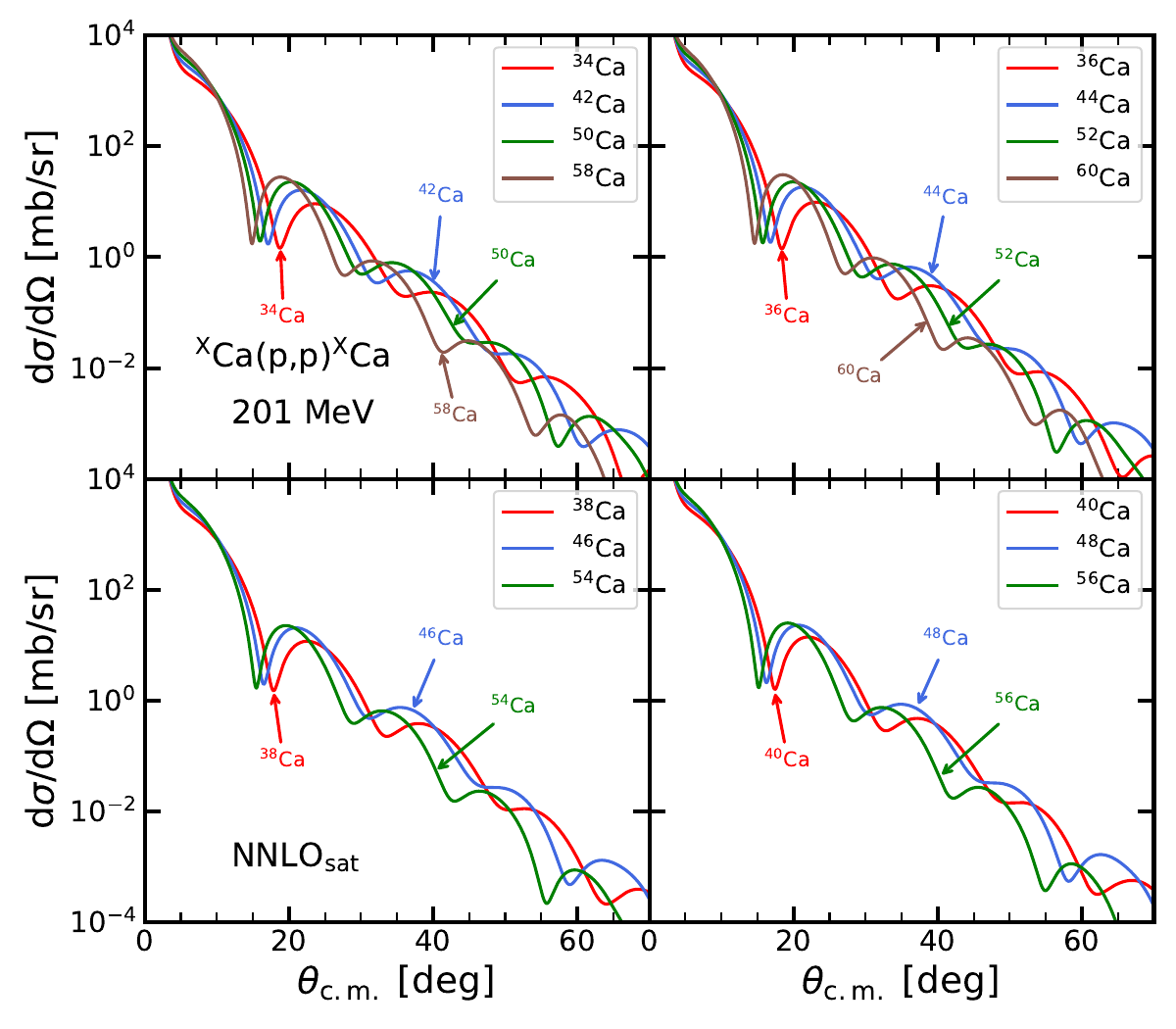}
\caption{\label{fig:Ca_chain_sig}
Differential cross section as a function of the c.m. scattering angle for proton elastic scattering off $^{\rm 36-60}$Ca isotopes at 201~MeV laboratory energy. The OPs are computed  with \sat{} and GkvADC(2) densities. The four panels show the trend of the cross section with increasing neutron-to proton asymmetry in the isotopic chain.  
}
\end{figure}

\begin{figure}[!h]
\includegraphics[width= \columnwidth]{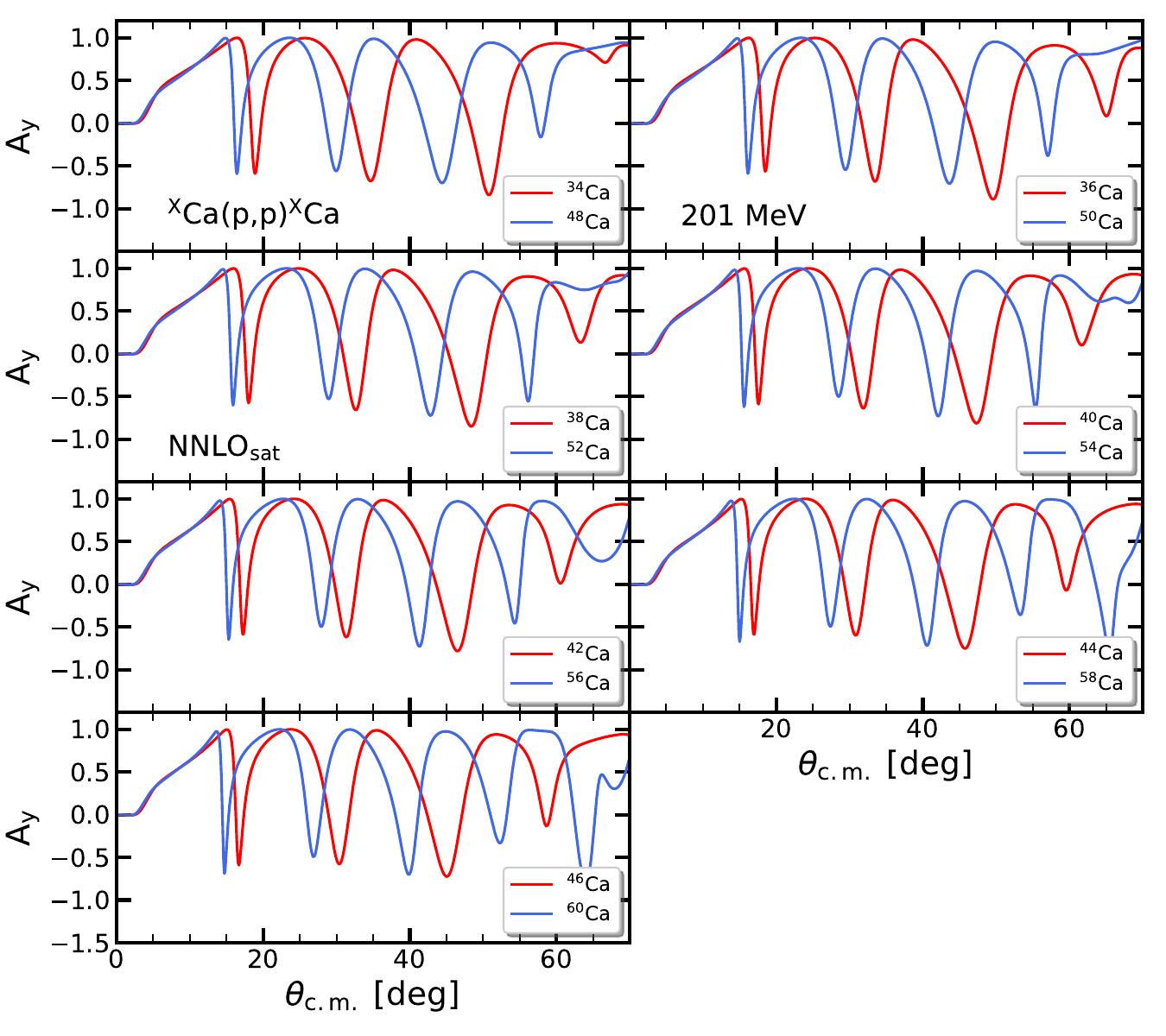}
\caption{\label{fig:Ca_chain_Ay} 
Analyzing power as a function of the c.m. scattering angle for proton elastic scattering off $^{\rm 36-60}$Ca isotopes at 201~MeV laboratory energy. The OPs are computed  with \sat{} and with GkvADC(2) densities. The four panels show the trend of the analyzing power with increasing neutron-to proton asymmetry in the isotopic chain.}
\end{figure}

\begin{figure}[t]
\includegraphics[width= \columnwidth]{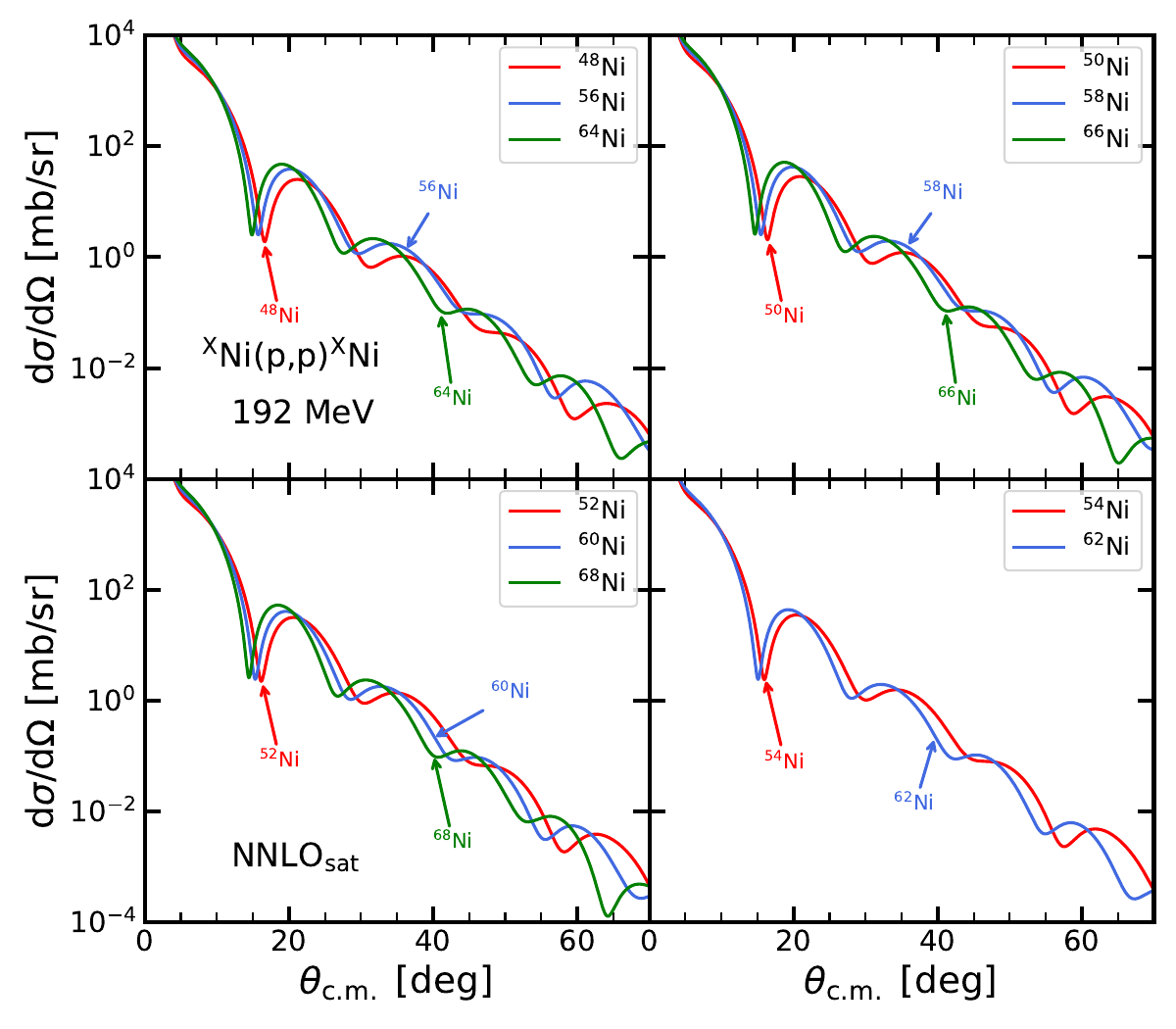}
\caption{\label{fig:Ni_chain_sig} 
Same as in Fig.~\ref{fig:Ca_chain_sig}
but  for proton elastic scattering off $^{\rm 36-60}$Ni isotopes at 192~MeV laboratory energy.}
\end{figure}

\begin{figure}[!h]
\includegraphics[width= \columnwidth]{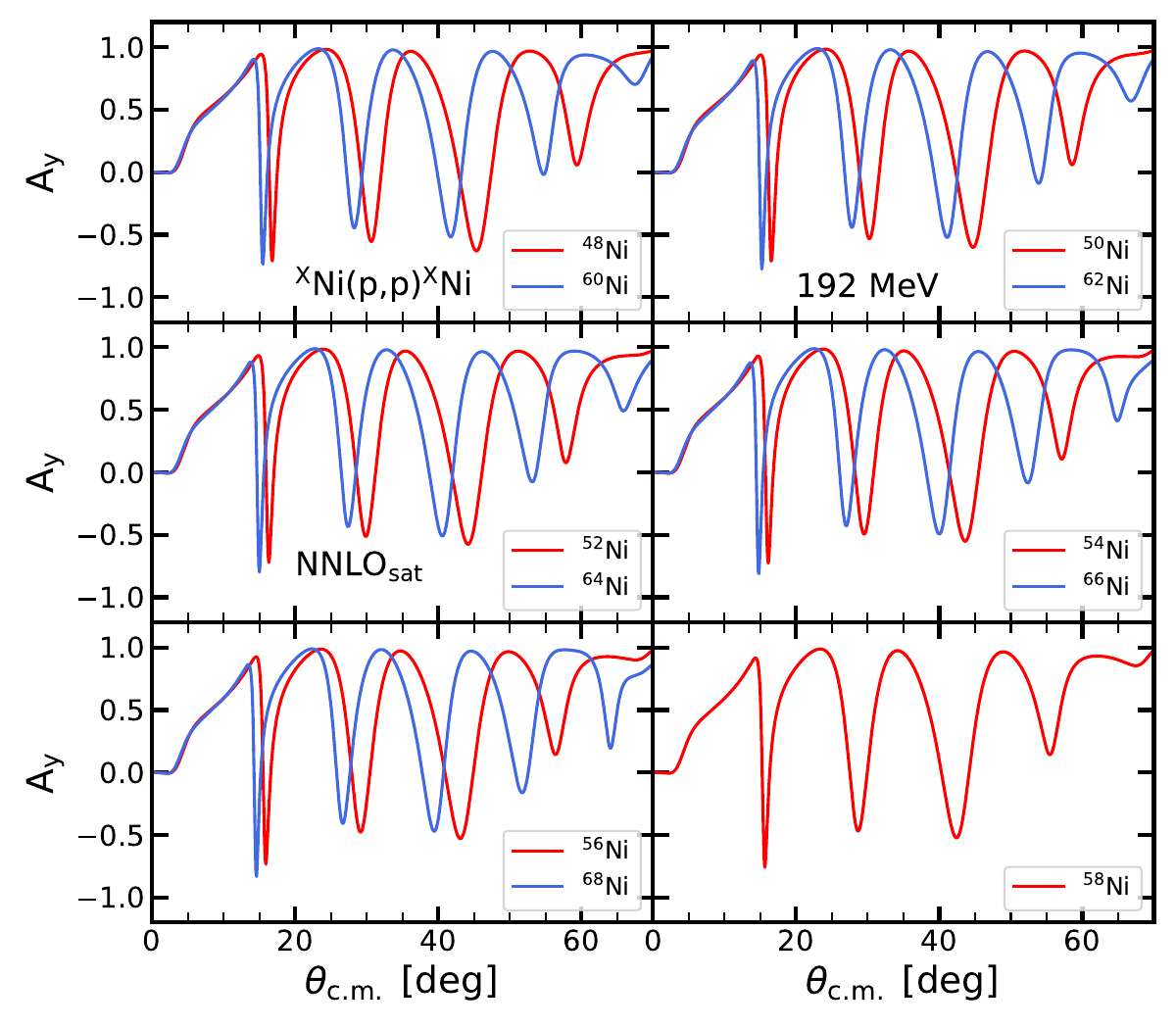}
\caption{\label{fig:Ni_chain_Ay} 
Same as in Fig.~\ref{fig:Ca_chain_Ay}
but  for proton elastic scattering off $^{\rm 36-60}$Ni isotopes at 192~MeV laboratory energy.}
\end{figure}

Given the generally satisfactory agreement between our microscopic OPs and experimental data, we produce theoretical predictions for elastic proton scattering off the whole Calcium and Nickel isotopic chains. These include cases for which experimental data are not yet available. We remind that the \sat{} interaction is known to predict the charge radii for the two $^{\rm 40,48}$Ca isotopes and the known data for the Ni chain~\cite{Soma2020lnl,Malbrunot2022prlNi} to good accuracy. However, it reproduces only partially the steep rise in Ca radii for $\mathrm{N} \geq 50$~\cite{Ruiz2016Nat_largeCa} while GkvADC(2) computations do not explain the bell shape for $20 < \mathrm{N} < 28$~\cite{Soma2020lnl}. These discrepancies are obviously transferred to the predicted scattering for these specific targets.
The results are displayed in Figs.~\ref{fig:Ca_chain_sig} and \ref{fig:Ca_chain_Ay} for $^{34-60}$Ca targets at 201 MeV and in Figs.\ref{fig:Ni_chain_sig} and \ref{fig:Ni_chain_Ay} for $^{48-68}$Ni at 192 MeV. We perform computations for the density matrix using GkvADC(2) and the \sat{} interaction in the $NN$ $t$ matrix. 
We  present here only a few  predictions for laboratory energies at about 200 MeV where the IA and our OPs model are fully justified, although similar calculations are easily doable at different energies.



The evolution of the differential cross sections and analyzing powers with increasing neutron-to-proton asymmetry is regular and similar for both isotopic chains.
The main effect of increasing the neutron number is a shift of the cross section minima toward smaller scattering angles. This compression of the diffraction minima is a direct consequence of the changes in the root mean square radius of the targets, which is included in phenomenological models by scaling the size of  Woods-Saxon potential as  $A^{1/3}$. In the \emph{ab initio} framework, the isotopic shifts of radii are computed microscopically and they are a direct prediction of the Hamiltonian. Hence, one has in principle the capability of making predictions beyond simple phenomenological trends. For our calculations, the limits of \sat{} mentioned above apply~\cite{Soma2020lnl,Malbrunot2022prlNi,Ruiz2016Nat_largeCa}.
Figs.~\ref{fig:Ca_chain_sig} and \ref{fig:Ca_chain_Ay} show that the compression of diffraction minima is similar for both Ca and Ni chains and the shift becomes stronger at large scattering angles.
This is clearly seen from the positions of the diffraction minima that shift toward smaller scattering angles. This is in general accompanied by a simultaneous increase in the height of the maxima. This behavior is similar for both Calcium and Nickel isotopic chains. Also for the analyzing powers, with increasing of the neutron-to-proton ratio, the main effect for both isotopic chains is a shift toward smaller scattering angles that increases with the scattering angles.

\section{Conclusions}
\label{Sect4}

We reported on a new advancement in the theory of microscopic OPs. This is part of an ongoing project, which aims at devising a framework for nucleon-nucleus elastic scattering that is free from phenomenology and sufficiently reliable to guide future experimental researches. Our OP is derived within the Watson multiple scattering theory, using $NN$ and $3N$ chiral interactions as only input. 
The final expression for the OP is obtained at the first order of the spectator expansion as the folding integral between the density of the target nucleus and the $NN$ $t$ matrix, representing respectively the structure and the dynamic part of the OP.

Earlier applications of this approach were affected by a lack of consistency between the calculation of the $t$ matrix and the one-nucleon density profile of the target, which were obtained with different techniques. This problem was recently overcome for light isotopes using the \emph{ab initio} NCSM, which can provide accurate descriptions of the target density adopting the same chiral interaction used in the computation of the $NN$ $t$ matrix. 
Within this framework, the model was then extended to describe antiproton scattering, to investigate the impact of $3N$ interactions on the dynamic part of the OP, and to be applied to nonzero spin targets~\cite{Vorabbi:2019ciy,Vorabbi:2020cgf,Vorabbi:2021kho}.
In general, however, the study of systems far from stability requires knowing microscopic OPs for a wide range of target isotopes, with medium and heavy mass, that are beyond the reach of foreseeable NCSM applications.

In this work, we have begun to exploit SCGF theory in aid to the current spectator model framework.
The SCGF approach presents better scaling of computational requirements with respect to the mass number that allows us to reach heavier systems, currently up to masses $\mathrm{A} \approx 140$, 
and provides fully nonlocal density matrices for the target. We presented and discussed results for differential cross sections and analyzing powers of elastic proton scattering up to 201 MeV in laboratory energy, where chiral interactions are still usable and one can rely on the impulse approximation.
Detailed computations for $^{40}$Ca and $^{66}$Ni targets demonstrate that the SCGF input is completely stable and scattering observables are well converged with respect to the model space, three-nucleon forces and many-body truncation already at the ADC(2) level.
We further compared to the available experimental data for elastic proton scattering off $^{40,48}$Ca and $^{58,60,62}$Ni targets. In all cases we obtained a very good reproduction of the experimental differential cross section and a remarkable description of the analyzing power, where the minima are correctly reproduced. Agreement with the experiment remained satisfactory down to $E_{lab}$=65~MeV energies for $^{40}$Ca even if this is somewhat below the limits of validity expected for the impulse approximation.

The good agreement between our results and experimental data gives us confidence in the reliability of the theoretical OPs. Therefore, we computed predictions for  elastic scattering off the whole Ca and Ni isotopic chains to investigate the evolution of the differential cross section and analyzing power with the increasing asymmetry between the number of neutrons and protons. For both isotopic chains we observed a compression of minima in the differential cross section toward smaller scattering angles when increasing neutron-to-proton asymmetry, as due to the bigger root mean square radii. This shift is in general accompanied by a simultaneous increase in the height of the maxima. The same shift towards smaller angles is also observed in the analyzing power.

In our opinion, the combination of the spectator model and SCGF theories offers remarkable opportunities for the physics of radioactive beams, and in particular toward resolving the long-standing issue of the lack of consistency between structure and reactions in the interpretation of data. 
While the SCGF method has so far been applied to closed-shell or, through its Gorkov formulations, to semi-magic open shell isotopes, it is plausible that computations on deformed bases will soon become available at least at the ADC(2) level. In fact, this will open the possibility of devising first-principles optical potentials for essentially all isotopes in the lower part of the Segr\`e chart. The SCGF approach can also be employed for providing two-nucleon spectral densities~\cite{Barbieri2004ee1pN}, which are the basis for extending the current model for the OP to the next term of the spectator expansion in two ways: accounting for the $3N$ force between the projectile and in-medium pairs and computing double-scattering events. At laboratory scattering energies below 100~MeV,  where the impulse approximation may become more questionable, the self-energy computed through SCGF theory is itself a viable \emph{ab initio} OP~\cite{PhysRevLett.123.092501}. Future work will be devoted to advancements along these lines.

\begin{acknowledgments}
The density matrices and OPs presented in this work are available in numerical format upon request to the authors.
The authors are grateful to Petr Navrátil for providing matrix elements of the employed interactions and for several useful discussions. 
This work is supported by the UK Science and Technology Facilities Council (STFC) through grants ST/Y000099/1, ST/P005314/1, and ST/L005516/1.
Calculations were performed by using HPC resources at the DiRAC DiAL system at the University of Leicester, UK, (BIS National E-infrastructure Capital Grant No. ST/K000373/1 and STFC Grant No. ST/K0003259/1), at the the National Energy Research Scientific Computing Center, a DOE Office of Science User Facility supported by the Office of Science of the U.S. Department of Energy under Contract No. DE-AC02-05CH11231, using NERSC award NP-ERCAP0020946, and from GENCI-TGCC, France, (Contracts No. A0130513012 and A0150513012).

\end{acknowledgments}

\newpage
\appendix
\section{Proton scattering observables}\label{AppX}

\begin{figure}[!b]
\includegraphics[width=\columnwidth]{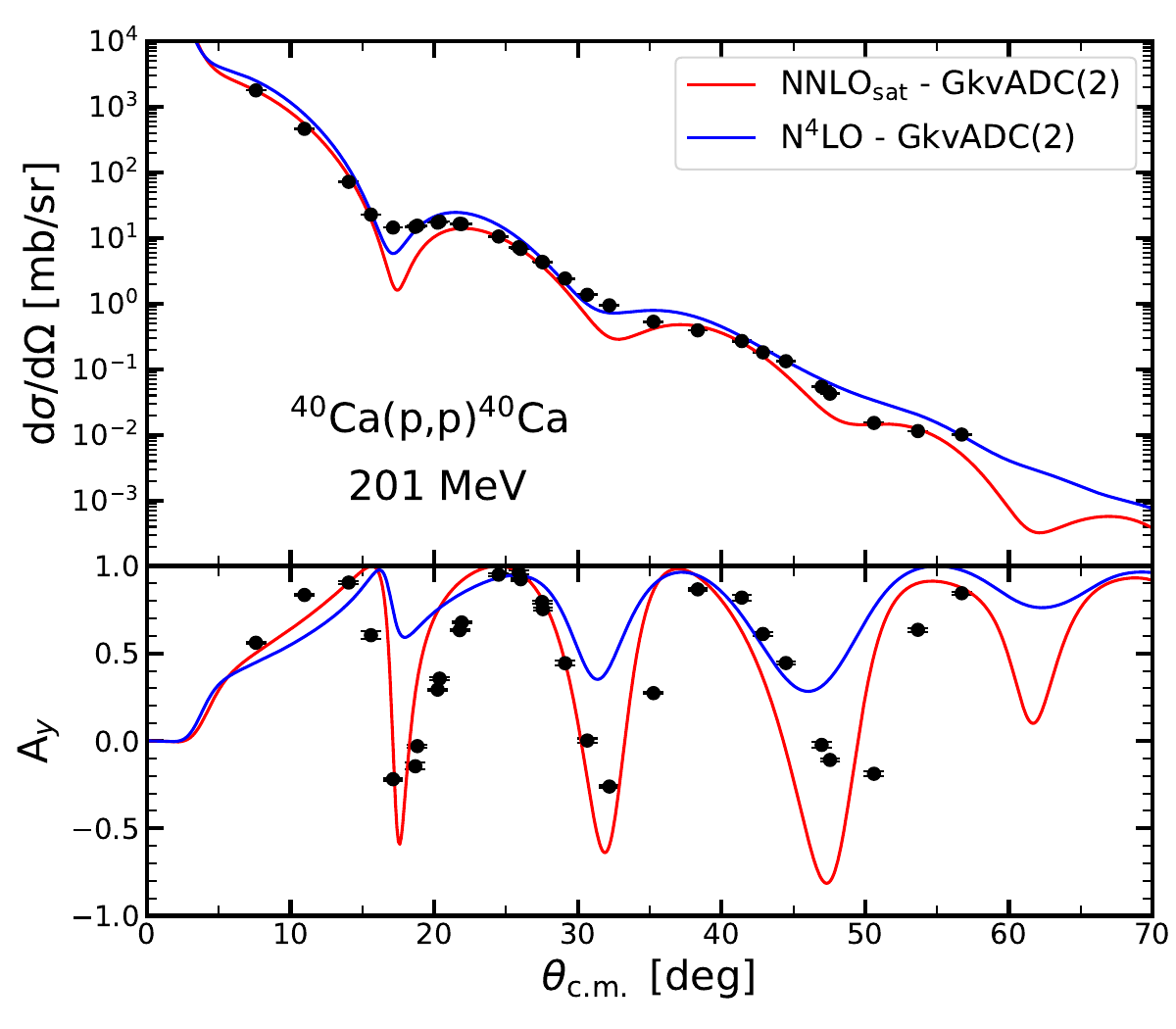}
\caption{\label{fig:Caxx_vs_exp1} 
Differential cross section (top panel) and analyzing power (bottom panel) as a function of the c.m. scattering angle $\theta_{\rm c.m.}$ for elastic proton scattering off $^{40}$Ca at a  laboratory energy of 201 MeV. The experimental data \cite{PhysRevC.47.1615} are compared with the results of microscopic OPs obtained using \sat{} (red curves) and N$^4$LO (blue curves) chiral interactions in the $NN$ $t$ matrix. In both cases, the nuclear density is obtained from GkvADC(2) SCGF calculations computed with \sat{}  with $N_{\rm max}$=13 and $\hbar\Omega$=14~MeV. }
\end{figure}

\begin{figure}
\includegraphics[width=\columnwidth]{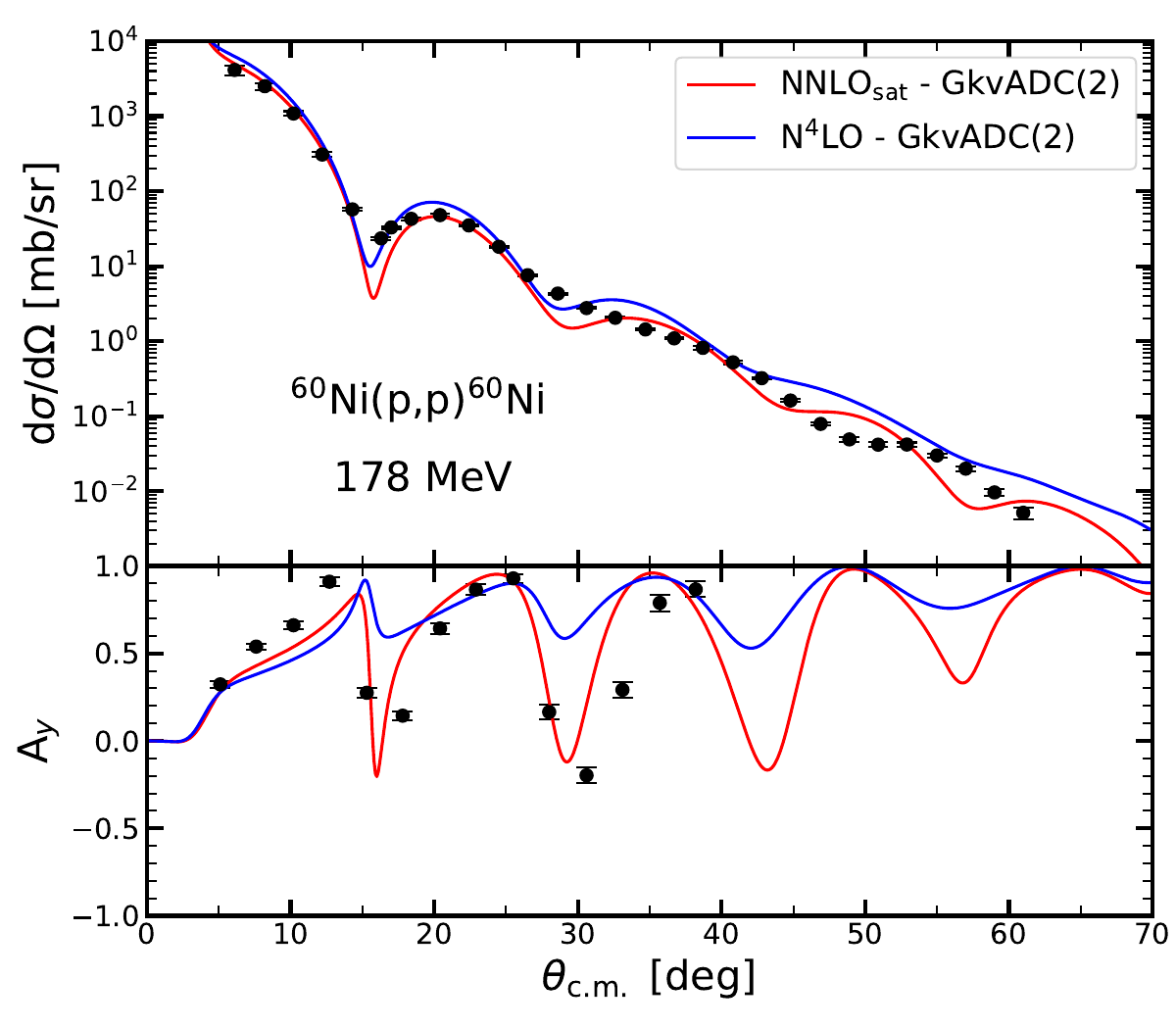}
\caption{\label{fig:Ni_vs_exp2}
Same as in Fig.~\ref{fig:Caxx_vs_exp1} but for elastic proton scattering proton off $^{60}$Ni at a laboratory energy of 178~MeV. Experimental data from Ref. \cite{INGEMARSSON1981426}.} 
\end{figure}

\begin{figure}
\includegraphics[width=\columnwidth]{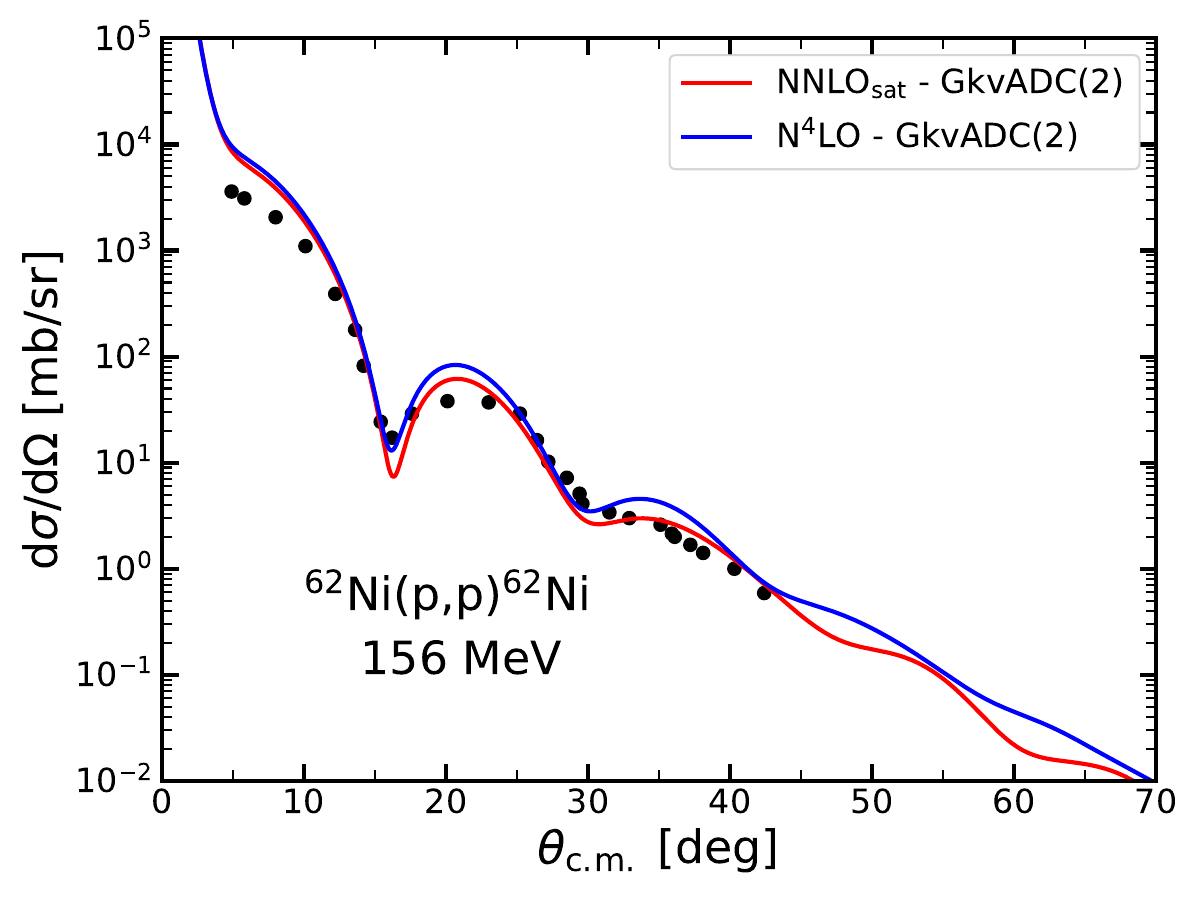}
\caption{\label{fig:Ni_vs_exp3}
Differential cross section as a function of the c.m. scattering angle for elastic proton scattering off $^{\rm 62}$Ni at 156~MeV laboratory energy. Line convention as in Fig.~\ref{fig:Caxx_vs_exp1}. Experimental data from Ref.~\cite{COMPARAT1974403}.}
\end{figure}

In supplement to the discussion of Sec.~\ref{Sec3B}, we show comparisons of our microscopic OP to experimental elastic proton scattering observables for additional targets and scattering energies.
Fig.~\ref{fig:Caxx_vs_exp1} displays differential cross section and analysing powers for $^{40}$Ca at 201~MeV laboratory energy, Fig.~\ref{fig:Ni_vs_exp2} is for $^{60}$Ni at 178~MeV, and Fig.~\ref{fig:Ni_vs_exp3} compares to the differential cross section of $^{62}$Ni at 156~MeV.
Overall, the findings of Sec.~\ref{Sec3B} in regards to the accuracy of prediction  employing the chiral interactions remain valid for all cases we tested and in the energy range 150-200~MeV, where our assumptions in deriving the OP are most reliable.

\newpage

\end{document}